\documentclass[AMA,Times1COL]{WileyNJDv5} 
\articletype{Article Type}
\received{Date Month Year}
\revised{Date Month Year}
\accepted{Date Month Year}
\journal{Journal}
\volume{00}
\copyyear{2023}
\startpage{1}

\usepackage{xcolor}

\begin{document}

\title{Assessing engineering wake models against operational data: insights from the Lillgrund wind farm wake steering campaign}

\author[1,4]{Diego Siguenza-Alvarado}
\author[2]{Matthew Harrison}
\author[3]{Mohammadreza Mohammadi}
\author[2]{Pragya Vishwakarma}
\author[2]{Ervin Bossanyi}
\author[2]{Lars Landberg}
\author[1]{Majid Bastankhah}

\authormark{SIGUENZA-ALVARADO \textsc{et al}}
\titlemark{ASSESSING ENGINEERING WAKE MODELS AGAINST OPERATIONAL DATA: INSIGHTS FROM THE LILLGRUND WIND FARM WAKE STEERING CAMPAIGN}

\address[1]{\orgdiv{Department of Engineering}, \orgname{Durham University}, \orgaddress{\state{Durham}, \country{UK}}}

\address[2]{\orgdiv{ Group Research \& Development}, \orgname{DNV}, \orgaddress{\state{Bristol}, \country{UK}}}

\address[3]{\orgdiv{
School of Science, Engineering, and Design}, \orgname{Teesside University}, \orgaddress{\state{Middlesbrough}, \country{UK}}}

\address[4]{\orgdiv{
Centro de Energías Renovables y Alternativas}, \orgname{Escuela Superior Politécnica del Litoral, ESPOL}, \orgaddress{\state{Campus Gustavo Galindo km. 30.5 Vía Perimetral, Guayaquil, 090902}, \country{Ecuador}}}

\corres{Corresponding author:\\ 
Majid Bastankhah \email{majid.bastankhah@durham.ac.uk}}



\abstract[Abstract]{Validating engineering wake models under real-world operational conditions is essential for improving wind farm performance predictions. This study utilises a unique dataset from the Lillgrund offshore wind farm, collected during the Horizon 2020 TotalControl project campaign, integrating synchronous Supervisory Control and Data Acquisition (SCADA) and Light Detection and Ranging (LiDAR) measurements under both baseline operation (i.e., no intentional yaw offset) and active wake steering scenarios. We assess four combinations of analytical wake models, each employing distinct formulations for velocity deficit, added turbulence, wake superposition and deflection, implemented in the LongSim modelling software developed by DNV. The analysis focuses on time-averaged wake velocity deficit profiles and turbine- and farm-wide power output, normalised by reference velocity and power. Model accuracy is quantified using mean absolute error (MAE) metrics. The evaluated models generally reproduce wake deficit trends under systematic variations in wake overlap in baseline cases, as well as wake deflection due to intentional yaw misalignment during the wake steering cases across a range of atmospheric conditions. Normalised velocity deficit MAE values range from 7\% to 15\%, with discrepancies primarily attributed to inflow heterogeneity, near-wake complexity and model-specific parameterisations. Power predictions reveal error accumulation with increasing farm depth. Model combinations incorporating cumulative wake superposition and refined turbulence schemes demonstrate improved agreement with field data; however, all models face challenges capturing localised flow features, with normalised turbine-level power output MAE ranging from 3\% to 23\%. Farm-wide power output errors ranged between -13\% and +30\%; though accurate farm-level predictions may mask compensating errors at individual turbines. Future studies should prioritise dynamic inflow characterisation and inclusion of blockage effects to enhance predictive reliability further.}

\keywords{analytical wakes, wake model validation, wind energy, Lillgrund wind farm}


\maketitle

\renewcommand\thefootnote{}

\renewcommand\thefootnote{\fnsymbol{footnote}}
\setcounter{footnote}{1}

\section{Introduction}\label{intro}

Wake effects within wind farms are critical to turbine performance because they reduce power output and increase fatigue loads. Owing to their computational efficiency, these effects are typically modelled using engineering-based analytical approaches. However, the accuracy in real atmospheric and operational conditions remains an open question, highlighting the need for rigorous field validation. The Lillgrund offshore wind farm, characterised by relatively dense turbine spacing and comprehensive Supervisory Control and Data Acquisition (SCADA) and Light Detection and Ranging (LiDAR) datasets, offers a valuable field-scale dataset for evaluating widely used engineering models, ranging from early formulations grounded in Reynolds-Averaged Navier–Stokes (RANS) theory \cite{ainslie1988calculating}, commonly adopted in industry, to more recent vortex-sheet models \cite{bastankhah2022vortex}.

Field validations of different analytical wake models have been reported in prior studies. Early assessments at Horns Rev revealed persistent discrepancies, with analytical models often underestimating wake losses \cite{barthelmie2009modelling}. These models captured general trends in wake width and power reduction, but significant uncertainties remained \cite{barthelmie2010quantifying}. Gaumond et al \cite{gaumond2014evaluation} suggested that such discrepancies may stem more from uncertainty in wind direction inputs than from inherent model limitations. Similarly, Göçmen et al \cite{goccmen2016wind} found improved model agreement with observations when accounting for wind direction variability. However, the absence of complete inflow data and limitations on dataset size underscored the need for high-resolution measurements to enable robust benchmarking, especially in tightly spaced farms like Lillgrund. Archer et al \cite{archer2018review} reported systematic overpredictions of power output, particularly for downstream turbines, while models performed better for upstream units. Hamilton et al \cite{hamilton2020comparison} further observed that model accuracy at Lillgrund deteriorates with farm depth and may be sensitive to parameter tuning. Doekemeijer et al \cite{doekemeijer2022comparison} evaluated the Gaussian-shaped wake model \cite{bastankhah2016experimental} using data from three offshore wind farms, finding generally good accuracy. However, they also noted a trade-off between turbulence intensity and wind direction variability, as well as reduced model performance for central turbines due to the effects of deep arrays and inflow heterogeneity. More recently, Bay et al \cite{bay2022addressing} improved model accuracy by incorporating near-wake effects \cite{blondel2020alternative} and cumulative wake interactions \cite{bastankhah2021analytical}, noting that accounting for flow acceleration could further enhance predictions.

The validation of engineering models becomes more challenging with the adoption of wake steering strategies, particularly under operational conditions where inflow variability cannot be fully controlled, which involve intentionally misaligning the yaw of turbines to deflect their wakes away from downwind turbines and reduce downstream power losses. Fleming et al~\cite{fleming2017field} compared model predictions with field measurements from two-turbine pairs, finding reasonable agreement in terms of increased power capture. A similar field experiment by Fleming et al~\cite{fleming2019initial,fleming2020continued} validated the Gaussian-Curl-Hybrid (GCH) model~\cite{bastankhah2016experimental, Martinez-Tossas2019, King2021}, showing good agreement with wake loss measurements and indicating that more advanced deep-array models may yield greater accuracy. Howland et al~\cite{howland2019wind} compared their model against long-term averaged field measurements from a six-turbine wind farm, again finding reasonable agreement during wake steering operations. Doekemeijer et al~\cite{doekemeijer2021field} conducted a field experiment involving wake steering on two- and three-turbine pairs at the Sedini onshore wind farm, where their model~\cite{bastankhah2016experimental} showed reasonable agreement with observed wake locations. However, significant discrepancies were observed in wake depth. Fleming et al~\cite{fleming2021experimental} applied the GCH model in a wake steering experiment involving two-turbine pairs within an eleven-turbine wind farm, finding that the model captures general wake loss trends but emphasised the need for validation in larger farms to account for factors such as blockage effects, secondary steering, wind farm boundary layers and deep-array interactions. Simley et al~\cite{simley2021results} also used the GCH model in a wake steering experiment at a seven-turbine commercial wind farm, finding reasonable agreement with field measurements for two-turbine interactions but noting underpredictions in energy gains for combined upstream and downstream turbines. More recently, blind tests conducted by Göçmen et al~\cite{gocmen_farmconners_2022} demonstrated that power gains from wake steering models exhibit high sensitivity to parameter choices, leading to divergent results even when using similar frameworks. These findings underscore the need for further validation studies before implementing wake control strategies.

Unlike earlier studies that validated engineering models using long-term field data (typically SCADA measurements averaged and directionally binned), we focus on specific operational and inflow conditions using 10-min averaged data. This analysis leverages the comprehensive Lillgrund dataset collected during the Horizon 2020 TotalControl campaign \cite{totalcontrol2021d1.1}, which includes SCADA data along with inflow and wake LiDAR measurements. Sood et al \cite{sood2022comparison} previously compared five cases from Lillgrund with 75-min averaged Large Eddy Simulations (LES) under baseline operation, reporting good overall agreement with measured wake and power outputs, although discrepancies were observed at individual turbines, particularly under aligned inflow conditions (wind direction roughly parallel to the turbine rows). These LES results are referenced here to provide contextual insight into model-data discrepancies at Lillgrund, rather than as a case-by-case benchmark against the analytical models evaluated in this study. In the present study, we evaluate both conventional and more recently developed steady-state analytical wake models implemented in LongSim, a modelling tool developed by DNV, and extend the analysis to a broader set of conditions encompassing both baseline operation and wake steering scenarios. We begin by examining variations in inflow direction, considering both aligned and non-aligned layouts and distinguishing between more aligned and more staggered wake-interaction configurations. Finally, we assess cases in which a turbine within the array is intentionally yawed clockwise or anticlockwise to evaluate the models' ability to represent wake deflection and the resulting impact on downstream power production.

The remainder of this paper is structured as follows. Section \ref{lillgrund} describes the Lillgrund offshore wind farm and outlines the field campaign, including the SCADA and LiDAR measurement setup, operational conditions, data processing and case selection criteria. Section \ref{models} presents the analytical wake models implemented in the LongSim software, detailing their respective treatments of wake deficit, turbulence, superposition, and deflection. In Section \ref{validation}, we describe the methodology used to validate model performance, including normalisation procedures and error metrics, and discuss the results under both baseline and wake steering conditions, focusing on the accuracy of wake velocity deficit and power output predictions. Finally, Section \ref{conclusions} summarises the key findings and provides recommendations for future model development and validation.

\section{Lillgrund Wind Farm}\label{lillgrund}

\subsection{Available Field Data}

\begin{figure*}[htbp]
\centerline{\includegraphics[width=0.95\textwidth]{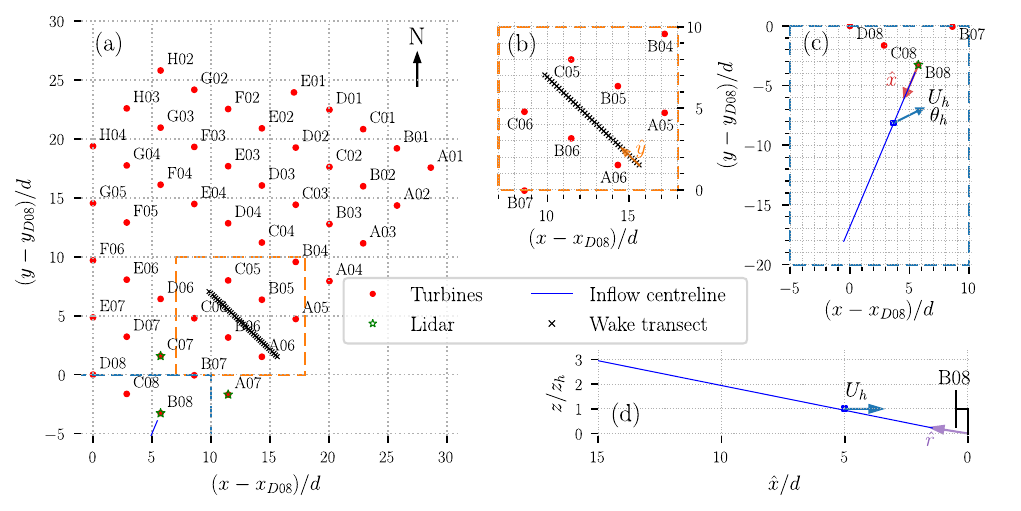}}
\caption{(a) Top-down view of the Lillgrund wind farm, showing the installed LiDAR scanners, relative to turbine D08 position. The LiDARs on turbines A07 and C07 measured the velocity within the wake transect synchronously, as represented by the black crosses and detailed in (b). The LiDAR on turbine B08 measured the inflow velocity, with its centreline indicated by the blue line and detailed in (c), alongside its side view in (d), where the square represents the hub-height measurement of horizontal velocity \(U_h\) and its direction \(\theta_h\). Adapted from Sood et al\cite{sood2022comparison}  \label{fig_layout}}
\end{figure*}

The Lillgrund wind farm comprises 48 Siemens SWT-2.3-93 turbines located approximately 10 km off the southern coast of Sweden. Arranged in a relatively compact layout, each turbine has a rated power \(\left(P_0\right)\) of 2.3 MW, a rotor diameter \(\left(d\right)\) of 93 m and a hub height \(\left(z_h\right)\) of 65 m. Wind farm flow and turbine operations data were collected during the measurement campaign of the Horizon 2020 TotalControl project\cite{totalcontrol2021d1.1}. Long-range LiDARs \cite{vasiljevic2016long} were used to measure inlet and wake flows, while SCADA turbine data were collected between August 2019 and February 2020. Figure \ref{fig_layout}a illustrates the wind farm layout and LiDAR positions. This section summarises the campaign-processed data reported by Sood et al\cite{sood2022comparison}, which should be consulted for details regarding the raw data processing.

The inflow LiDAR on turbine B08 performed repeating plan position indicator (PPI) sector scans with an azimuth sweep of 60\(^\circ\) and an elevation angle of 8\(^\circ\). The inflow centreline is depicted in Figure \ref{fig_layout}c,d, and any misalignments were corrected using the turbine’s nacelle direction signal, thereby minimising directional misalignment effects and helping to ensure that wind speed measurements remained representative of the inflow. The inflow data consist of a time series recorded every 30 s (0.033 Hz), containing processed wind speed and direction measurements for range gates spanning 70--1490 m in 20-m increments. The hub height measurement corresponds to a gate range of 430 m, located approximately \(5d\) upstream of B08. It is used as the wind farm reference velocity \(\left(U_h\right)\) and direction \(\left(\theta_h\right)\) in this study.
Additionally, two wake-scanning LiDARs, installed on turbines A07 and C07, conducted coordinated dual-Doppler trajectory scans to measure wake effects and intra-farm flows. These measurements were synchronised in both space and time across the transect shown in Figure \ref{fig_layout}b at three heights: 6, 70 and 130 m. The wake data used in this study consist of 10-min averaged 2D wind fields at a height of 70 m. No additional filtering or spatial smoothing was applied beyond the temporal averaging inherent to the processed dataset, and further details on the underlying measurement uncertainties are provided in Sood et al~\cite{sood2022comparison}.

The SCADA operational data from all 48 turbines in the wind farm includes active power, blade pitch angles (blades A, B and C), nacelle direction, nacelle wind speed and rotor speed. These signals were recorded at 2-s intervals (0.5 Hz). Due to inconsistencies in the raw nacelle direction signals across turbines, individual calibration was performed to ensure alignment, resulting in corrected values in the final dataset.

\begin{figure*}[htbp]
\centerline{\includegraphics[width=0.99\textwidth]{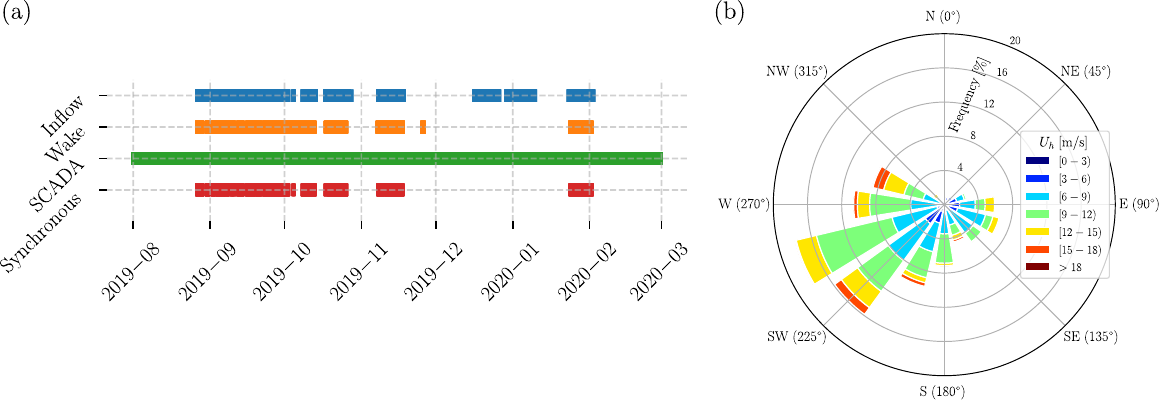}}
\caption{(a) Lillgrund data timeline in 10-min timestamps. (b) Wind rose depicting wind speed and direction at hub height, based on the inflow measurement LiDAR data during synchronous timestamps.  \label{fig_timeline_wind_rose}}
\end{figure*}

Figure \ref{fig_timeline_wind_rose}a shows the availability of data from various sources over different periods during the campaign. Our analysis focuses on timestamps labelled as 'synchronous' in Figure \ref{fig_timeline_wind_rose}a and highlighted in red where data from all sources are available. To identify this synchronous dataset, the inflow and SCADA data are resampled to 10-min averages to align with the wake dataset. The wind rose representation of the reference velocity \(U_h\) and direction \(\theta_h\) during the synchronous timestamps is shown in Figure \ref{fig_timeline_wind_rose}b.

\subsection{Atmospheric Conditions}

Accurate wake modelling depends on several inflow parameters, notably the mean wind speed, turbulence intensity, vertical shear, wind-direction veer, and atmospheric stability. Wind speed, shear and veer are obtained directly from the inflow LiDAR measurements and are discussed in detail in the case-selection analysis in Section~2.4. This subsection therefore focuses on the characterisation of turbulence intensity and atmospheric stability, as these quantities require additional processing beyond the standard LiDAR-derived variables.

\begin{figure*}[htbp]
\centerline{\includegraphics[width=0.95\textwidth]{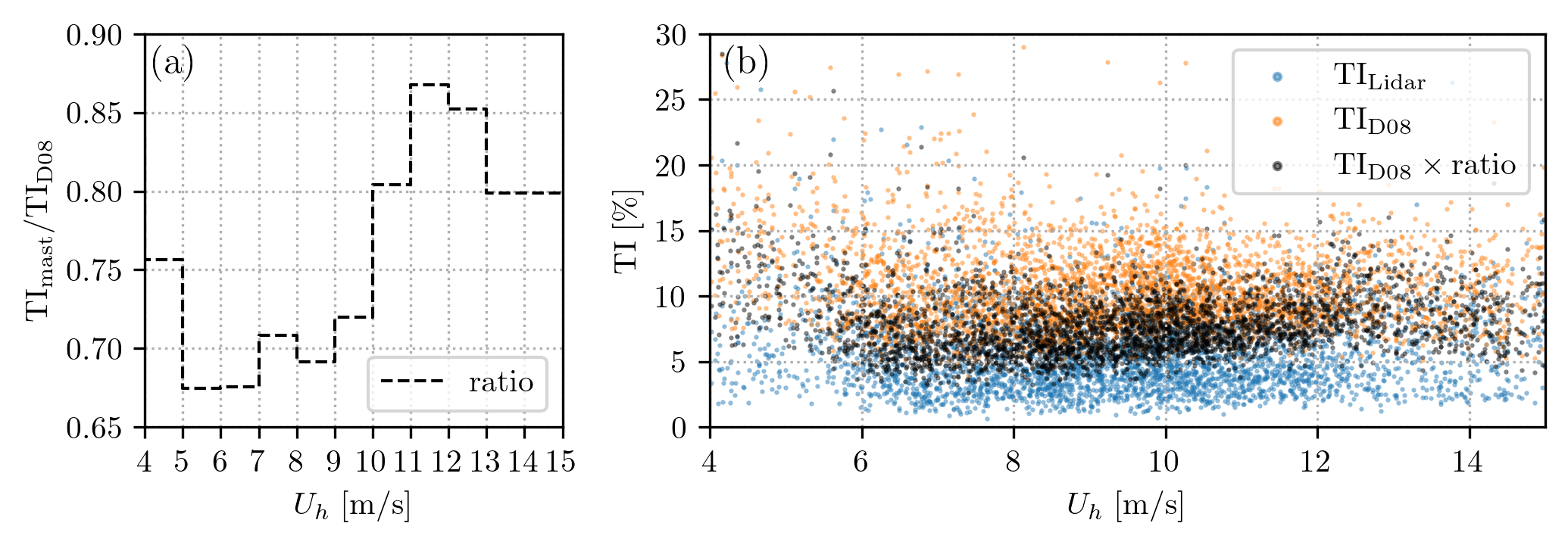}}
\caption{(a) The ratio of turbulence intensity (TI) measured with the met-mast to the turbine D08 nacelle's wind speed, grouped by wind speed bins of 1m/s, as reported by G\"{o}\c{c}men and Giebel \cite{goccmen2016estimation}. (b) TI from the synchronous timestamps, filtered for $4\textrm{m/s}<U_h<15\textrm{m/s}$ and $135^\circ<\theta_h<320^\circ$, measured using the inflow LiDAR and SCADA from the D08 nacelle wind speed; the black dots represent D08 nacelle TI multiplied by the ratio.     \label{fig_TI}}
\end{figure*}

Turbulence intensity (TI) characterises the variability of the incoming wind, plays a critical role in wake recovery and is therefore essential for accurate wake modelling. TI is defined as
\begin{equation}
\textrm{TI} = \frac{\sigma_u}{\overline{u}}, 
\label{eqTI}
\end{equation}
where $\overline{u}$ represents the time-averaged wind speed, and $\sigma_u$ denotes the standard deviation of the wind speed $u$ over a 10-min period. 
In the current field campaign, TI can be estimated using two primary sources: (i) LiDAR-based inflow measurements and (ii) nacelle wind speed data from the most upstream turbine relative to the prevailing wind direction, for example, turbine D08 during south-westerly winds. Each of these approaches presents specific limitations. The inflow LiDAR, which operates at a sampling frequency of 0.033~Hz, underestimates TI due to its limited ability to resolve small-scale turbulent fluctuations. In contrast, the nacelle signal, recorded at 0.5 Hz, captures finer-scale variability but tends to overestimate the inflow TI due to its location behind the rotor. To address this discrepancy, we refer to the study by G\"{o}\c{c}men and Giebel~\cite{goccmen2016estimation}, which provides a detailed comparative analysis of TI measurements at the Lillgrund offshore wind farm. Their dataset, collected using multiple instruments operating at 1 Hz between June 2012 and January 2013, includes both meteorological mast sensors positioned approximately 1.4 rotor diameters upstream of turbine D08 and nacelle-mounted sensors on the same turbine. In their analysis, the ratio between the mast-derived and nacelle-derived TI values, denoted as $\textrm{TI}_{\textrm{mast}}$ and $\textrm{TI}_{\textrm{D08}}$, respectively, was quantified for filtered wind speed and direction intervals (Figure~\ref{fig_TI}a). Based on this reference, we apply the same ratio to correct nacelle-based TI measurements in our field campaign. Figure~\ref{fig_TI}b presents the comparison of synchronous LiDAR and D08 nacelle signals collected during our campaign. By adjusting the nacelle-based measurements using the ratio derived from the G\"{o}\c{c}men and Giebel data, we estimate inflow TI values in the range of 5--10\%, which is consistent with mast-derived values reported in their study.

The atmospheric stability during the field campaign is characterised using the Monin--Obukhov length $(L)$. Direct temperature and surface heat-flux measurements were not performed as part of the measurement campaign; therefore, the ERA5 reanalysis dataset~\cite{hersbach2020era5} is used to estimate $L$. ERA5 data are sampled hourly from the nearest offshore grid point, located approximately 22~km southwest of the Lillgrund wind farm. These stability estimates are used to characterise the prevailing atmospheric conditions during the campaign and are employed in the modelling only where explicitly noted in subsequent sections.

\subsection{Turbine Operational Features}

Turbine operational characteristics influence wake behaviour and must therefore be represented consistently within the wake-modelling framework. In addition to geometric parameters such as rotor diameter and hub height, which affect wake expansion and interaction with the atmospheric boundary layer, turbine aerodynamic operating states determine the magnitude of momentum extraction and thus the strength of the generated wake. In particular, the blade-averaged pitch angle $(\beta)$ directly affects the thrust coefficient and consequently the downstream wake evolution.

\begin{figure*}[htbp]
\centerline{\includegraphics[width=0.95\textwidth]{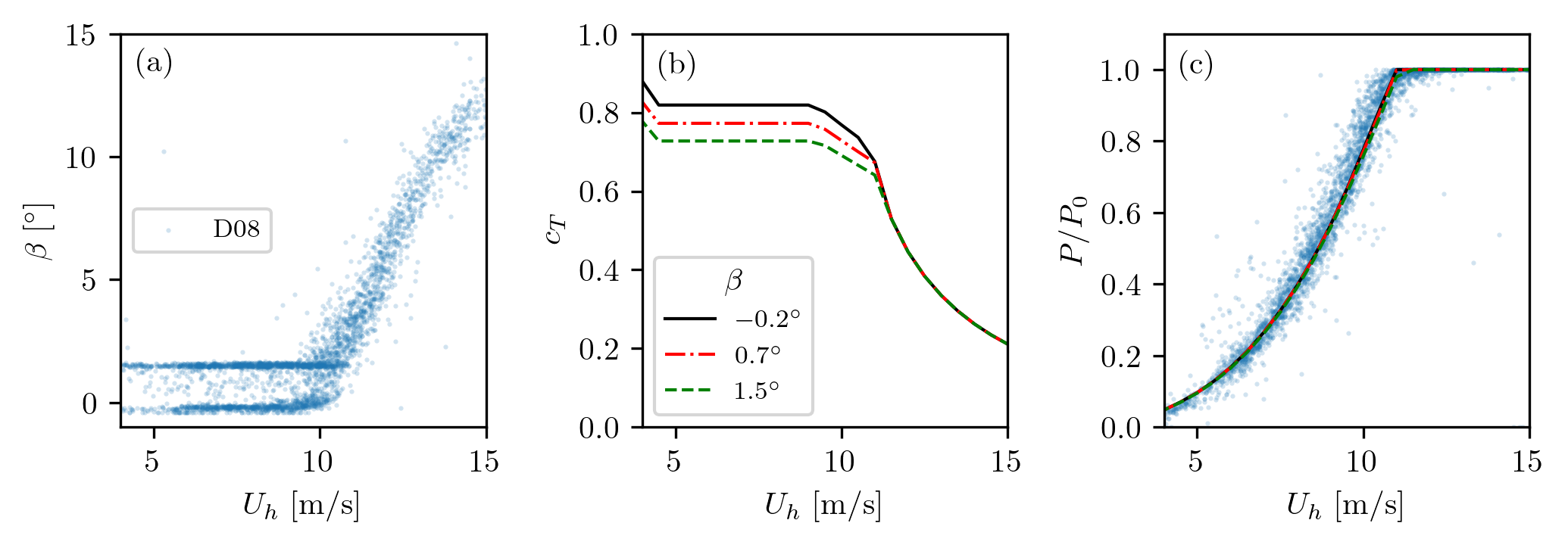}}
\caption{(a) Averaged pitch angle of turbine D08 from the filtered synchronous timestamps. (b) Thrust coefficient curves of modelled non-yawed turbines with representative pitch angles. (c) Normalised power curves compared against D08 field data. The legends in (a) and (b) also apply to (c).      \label{fig_turbine}}
\end{figure*}

In the analysis of the Lillgrund field campaign, Sood et al~\cite{sood2022comparison} modelled the Siemens SWT-2.3-93 turbines assuming a constant zero pitch angle. Examination of the SCADA data, however, shows that most turbines operated with non-zero pitch angles even at below-rated wind speeds. Figure~\ref{fig_turbine}a illustrates the averaged pitch angle of turbine D08 for the filtered synchronous timestamps, revealing two distinct operating trends at approximately $\beta=-0.2^\circ$ and $\beta=1.5^\circ$ under below-rated conditions. Similar behaviour is observed for other upstream turbines, whereas turbines operating within wakes predominantly exhibit pitch angles around $\beta=-0.2^\circ$, potentially reflecting intentional derating strategies employed by the wind-farm operator.

To account for the influence of these non-zero pitch angles on turbine performance, thrust and power coefficients are obtained using DNV~Bladed. Bladed is an aero-servo-hydro-elastic simulation tool for wind turbines that employs a blade element momentum (BEM) formulation for rotor aerodynamics, coupled with a multi-body structural dynamics model and a high-frequency controller interface~\cite{bladed}. To derive thrust and power curves suitable for input to the engineering wake models, simulations are performed under steady, uniform inflow conditions. Rotor tilt is set to zero, and effects of turbulence, vertical shear and upflow are intentionally neglected at this stage, such that the resulting thrust and power coefficients represent mean, rotor-averaged quantities as functions of wind speed, pitch angle and yaw misalignment.

The effects of turbulence and shear on wake development are subsequently accounted for at the wake-modelling level within LongSim (see Section~\ref{models}). Several thrust and power curves are generated for representative pitch angles, with $\beta=0.7^\circ$ included for specific cases. In addition, yaw-dependent performance curves are derived for yaw misalignment angles ranging from $\gamma=-30^\circ$ to $\gamma=30^\circ$ in $5^\circ$ increments, enabling consistent treatment of wake steering effects through prescribed turbine operating coefficients.

Figure~\ref{fig_turbine}b shows the resulting thrust coefficient curves for different pitch angles at zero yaw, highlighting noticeable sensitivity at below-rated wind speeds. In contrast, the corresponding power curves shown in Figure~\ref{fig_turbine}c collapse closely across the considered pitch angles. These modelled power curves are compared with field measurements from turbine D08 for the filtered synchronous timestamps, demonstrating good agreement in the mean sense despite the variability in the SCADA data.

\subsection{Case selection}

To evaluate the analytical wake models, we selected field data with an emphasis on data quality and directional coherence across the array. Our analysis focused on seven baseline cases (i.e., no intentional yaw misalignment) and seven wake steering cases, which were subsequently compared to the predictions of the analytical wake models. In addition to removing timestamps that have missing information, those with high yaw angles were also excluded to improve directional coherence across the array and reduce unintended yaw misalignment effects, without imposing constraints on turbulence intensity, shear or atmospheric stability. The specific turbine yaw angle \((\gamma_i)\) was calculated as follows:  
\begin{equation}  
\gamma_i = \theta_h - \theta_i,  
\label{eq_gamma}  
\end{equation}  
where \(\theta_i\) is the SCADA signal of the nacelle wind direction and the subscript \(i\) represents the turbine label, shown in Figure~\ref{fig_layout}a. The reference wind direction, \(\theta_h\), is obtained from the inflow LiDAR measurements. Note that a positive \(\gamma_i\) denotes an anticlockwise yaw from a top-down view. By minimising the average of the absolute yaw angles \((\Gamma)\), defined as 
\begin{equation}  
\Gamma = \frac{1}{n}\sum_{i}|\gamma_i|,  
\label{eq_Gamma}  
\end{equation}
where $n$ is the number of turbines considered in the average, we aim to reduce unintended yaw misalignment across the array and improve directional coherence for model-data comparison. It is important to note that turbine A02 exhibited outlier yaw angles and was excluded from the averaging. Additional filtering was also applied to ensure turbines that are located upstream of our measurement domain were in operation. For instance, timestamps with negative active power for turbine D08 and those with reference inflow velocities below cut-in speeds were excluded.  

\subsubsection{Baseline Cases (No Intentional Yaw Offset)}

The baseline cases are intended to characterise model behaviour under the operating conditions captured in the available non-steering measurements and are not treated as a controlled reference for comparison with the wake-steering cases.

The selection criteria for the baseline cases (denoted by the subscript \(0\) hereafter) were based on the reference wind direction \(\theta_h\) (with a tolerance of \(\pm 1^\circ\)) and the minimum \(\Gamma\). Beginning with an aligned wind farm, case \(a_0\) occurs at \(\theta_h \approx 222^\circ\), which approximates to the most likely wind direction according to the wind rose in Figure~\ref{fig_timeline_wind_rose}b and results in an aligned wind farm configuration with full wake conditions, characterised by maximum downstream wake overlap. Subsequent cases systematically increase the reference wind direction by \(5^\circ\) to sample configurations with varying degrees of wake overlap, labelling the cases alphabetically until reaching case \(g_0\) at \(\theta_h \approx 252^\circ\), where the wind farm is nearly aligned once more. Overall, the baseline cases show \(\Gamma\) ranging from \(1.50^\circ\) for case \(\textrm{e}_0\) to \(3.52^\circ\) for case \(\textrm{b}_0\).

\begin{figure*}[htbp]
\centerline{\includegraphics[width=0.95\textwidth]{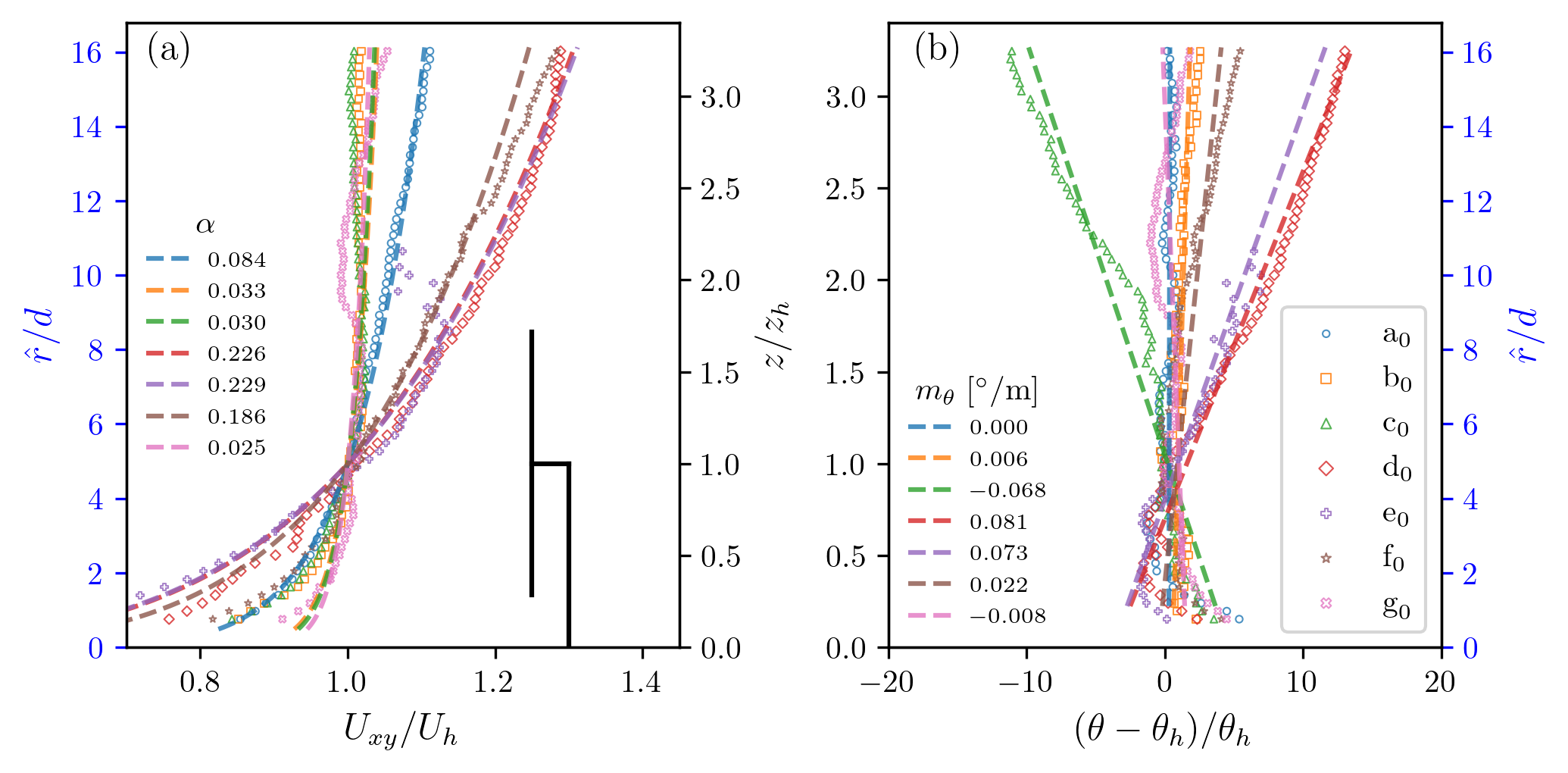}}
\caption{Baseline cases (from $\textrm{a}_0$ to $\textrm{g}_0$) inflow conditions: (a) Normalised horizontal velocity profile along its range gate, where the dashed lines represent the power-law fitting. (b) Normalised wind veer profile along its range gate, where the dashed lines represent its linear fitting. The framed legend in (b) applies to both subfigures. \label{fig_inflow_0}}
\end{figure*}

The inflow shear in the seven baseline cases, obtained from inflow LiDAR measurements, can be seen in Figure~\ref{fig_inflow_0}a. The scatter points represent the normalised horizontal velocity \((U_{xy})\) measured by the inflow LiDAR along its range gate (see the direction \(\hat{r}\) in Figure~\ref{fig_layout}d). Note that data are missing in case \(e_0\) above the top tip rotor height, as indicated by the turbine schematic in Figure~\ref{fig_inflow_0}a. For the engineering modelling performed later in Section \ref{lillgrund}, we assume incoming velocity profiles follow a power law, and the observations were fitted to determine the shear exponents \(\alpha\) for each case, as depicted in Figure~\ref{fig_inflow_0}a.  It is worth noting that, as mentioned in Section \ref{lillgrund}.1, the inflow LiDAR measurements are taken along a line at an elevated angle. In this analysis, we neglect any potential heterogeneity in inflow and assume that the results can be projected onto a vertical profile. The vertical variation of the wind direction, termed wind veer, is also evident in different cases as shown in Figure~\ref{fig_inflow_0}b. The veer slope, \(m_{\theta}\), is defined as the linear fit to the data, and its value for different cases is shown in Figure~\ref{fig_inflow_0}b. Except for cases \(a_0\) and \(c_0\), the veer slope is positive, indicating that the wind direction changes clockwise with height, which is typically expected in the Northern Hemisphere. To ensure a consistent description of inflow across all cases, notably when several upper LiDAR range gates were missing or unreliable, the vertical profiles of wind speed and direction were fitted using the power-law and linear-veer formulations introduced below, avoiding mixing fully measured and partially reconstructed inflow profiles in the model evaluation. Following the observations in Figure \ref{fig_inflow_0}, for each case, we assume a vertical inflow velocity profile with a streamwise velocity (aligned with \(\theta_h\) at the hub height), \(U_0\), defined by
\begin{equation}
U_0(z) = U_{h}(z/z_h)^\alpha \cos(m_\theta (z-z_h)), 
\label{eq_U}
\end{equation}
and a lateral velocity, \(V_0\), defined by:
\begin{equation}
V_0 = U_{h}(z/z_h)^\alpha \sin(m_\theta (z-z_h)). 
\label{eq_V}
\end{equation}

Table~\ref{tab1} highlights atmospheric conditions for the seven baseline cases. For example, the wind speed, \(U_h\), varies between \(6.66\) and \(9.05~\textrm{m/s}\), corresponding to below-rated speeds, with a turbulence intensity (TI) between \(5.18\%\) and \(8.81\%\). Additionally, the Monin-Obukhov length, \(L\), encompasses a range of stability conditions, with a tendency towards weakly stratified or neutral stability classes, according to the classification shown by Ruisi and Bossanyi \cite{ruisi2019engineering}.

\begin{table*}[!t]%
\centering %
\caption{Baseline cases' model parameters.\label{tab2}}%
\begin{tabular*}{\textwidth}{@{\extracolsep\fill}llllllll@{\extracolsep\fill}}
\toprule
\textbf{Case} & \textbf{Time} (YYYY-MM-DD HH:MM)  & $\theta_h~(^\circ)$  & $U_h~(\textrm{m/s})$  & TI $(\%)$ & $\alpha$ & $m_{\theta}~(^\circ/\textrm{m})$ &  $L~(\textrm{m})$\\
\midrule
$\textrm{a}_0$ & 2019-10-23 10:00  & 222.74 & 7.70  & 5.18  & 0.084 & 0.000  & -236 \\
$\textrm{b}_0$ & 2019-11-08 05:40  & 226.79  & 8.04  & 5.36  & 0.033 & 0.006  & 1221 \\
$\textrm{c}_0$ & 2019-11-08 08:50  & 231.89  & 6.66 & 7.37  & 0.030 & -0.068  & -554 \\
$\textrm{d}_0$ & 2019-10-25 16:10  & 236.31 & 8.65  & 6.31  & 0.226 & 0.081  & 417 \\
$\textrm{e}_0$ & 2020-01-31 16:50  & 242.30 & 9.05  & 6.40  & 0.229 & 0.051  & 243 \\
$\textrm{f}_0$ & 2019-10-22 19:00  & 246.15 & 8.85  & 5.94  & 0.186 & 0.022  & 490 \\
$\textrm{g}_0$ & 2019-10-10 09:30  & 251.05 & 8.41  & 8.81  & 0.025 & -0.008  & -224 \\

\bottomrule
\end{tabular*}

\end{table*}

\subsubsection{Wake Steering Cases}

\begin{figure*}[htbp]
\centerline{\includegraphics[width=0.95\textwidth]{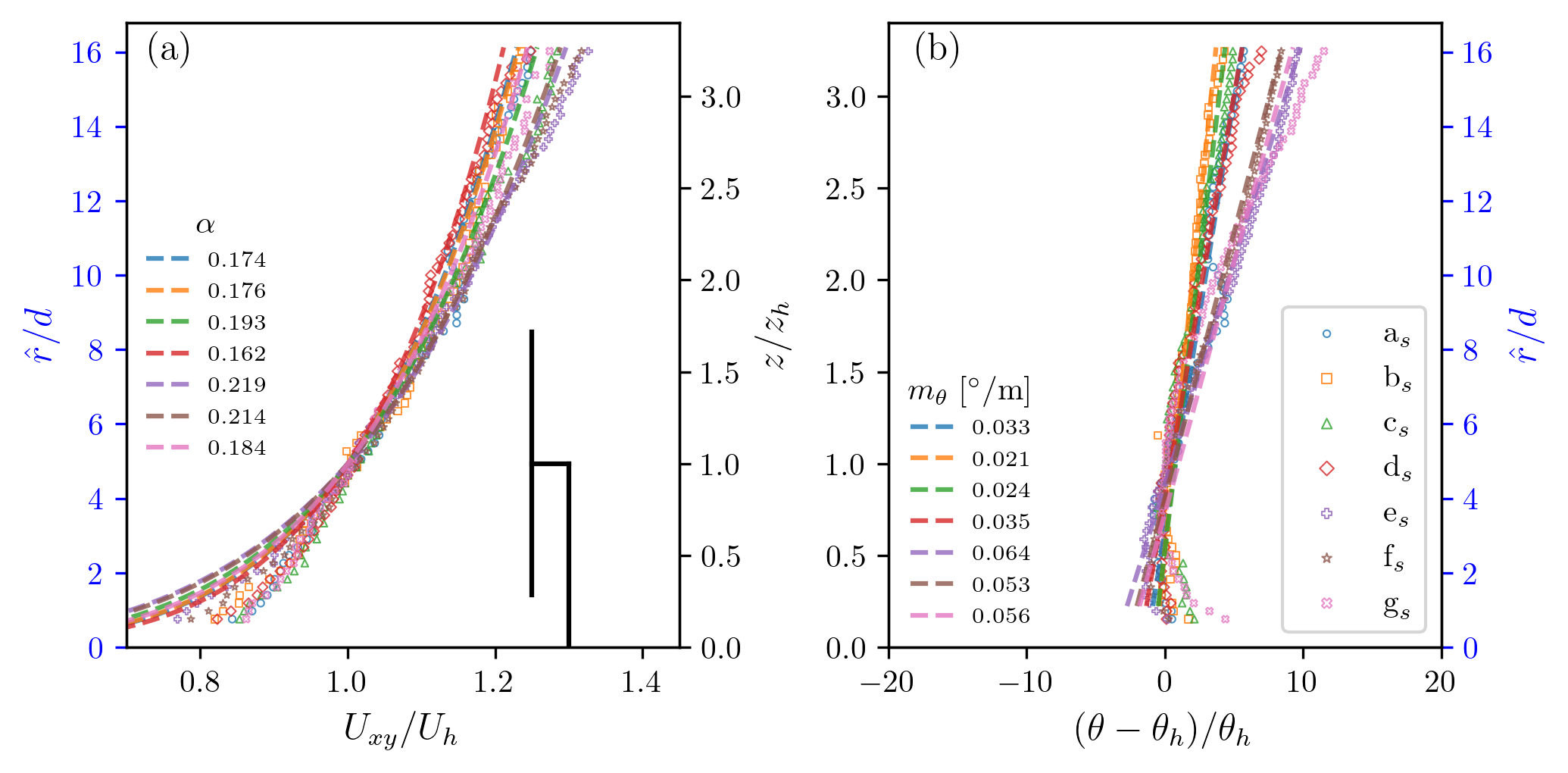}}
\caption{Steering cases (from $\textrm{a}_s$ to $\textrm{g}_s$) inflow conditions: (a) Normalised horizontal velocity profile along its range gate, where the dashed lines represent the power-law fitting. (b) Normalised wind veer profile along its range gate, where the dashed lines represent its linear fitting. The framed legend in (b) applies to both subfigures. \label{fig_inflow_s}}
\end{figure*}

The wake-steering campaign at Lillgrund took place over three nights: January 27, 28 and 29, 2020. The internal turbine B06 was tested for approximately 4--6 h each evening, alternating between positive, zero and negative yaw angles. For the wake steering cases (denoted by the subscript \(s\) and called steering cases hereafter for brevity), the selection criteria include a relatively low \(\Gamma\) of \(\approx 2^\circ\), excluding \(\gamma_{\mathrm{B06}}\) from the average. Additionally, we chose cases with clockwise or anticlockwise B06 yaw angles greater than \(15^\circ\), which result in noticeable wake deflection.  Moreover, we ensure that the yaw angle of turbines (A06, C06 and D06) immediately upstream of the wake transect are recorded in the SCADA data for all chosen cases. Wake steering cases were collected outside daytime hours and differ systematically from the baseline cases in wind-speed regime and atmospheric conditions. Accordingly, baseline and steering cases are not intended to be directly compared to isolate wake steering effects; instead, they are analysed independently to examine model sensitivity, consistency and limitations across distinct operating regimes. See the incoming velocity profiles in Figure~\ref{fig_inflow_s}a, where shear exponents vary between 0.174 and 0.184 and positive veer profiles in Figure~\ref{fig_inflow_s}b, with veer slopes ranging from \(0.033^\circ/\textrm{m}\) to \(0.056^\circ/\textrm{m}\). Atmospheric parameters for the seven steering cases are shown in Table~\ref{tab3}, where the cases are sorted alphabetically with increasing wind direction \(\theta_h\). There are four cases with anticlockwise and three with clockwise \(\gamma_{\mathrm{B06}}\). Most steering cases occur at wind speeds near or above rated conditions, except for case \(g_s\), while exhibiting TI values comparable to those of the baseline scenarios. The estimated Monin-Obukhov length \(L\) indicates predominantly weakly stable to neutral conditions across the steering cases, with no unstable conditions observed.

\begin{table*}[!t]%
\centering %
\caption{Steering cases model parameters.\label{tab3}}%
\begin{tabular*}{\textwidth}{@{\extracolsep\fill}lllllllll@{\extracolsep\fill}}
\toprule
\textbf{Case} & \textbf{Time} (YYYY-MM-DD HH:MM) & $\gamma_{\mathrm{B06}}~(^\circ)$   & $\theta_h~(^\circ)$  & $U_h~(\textrm{m/s})$  & TI $(\%)$ & $\alpha$ & $m_{\theta}~(^\circ/\textrm{m})$ &  $L~(\textrm{m})$\\
\midrule
$\textrm{a}_s$ & 2020-01-29 23:10 & -24.00 & 251.59 & 12.42  & 8.82  & 0.174 & 0.033  & 746 \\
$\textrm{b}_s$ & 2020-01-29 22:30 & 19.04 & 252.41  & 12.01  & 7.79  & 0.176 & 0.021  & 573 \\
$\textrm{c}_s$ & 2020-01-29 22:20 & 22.58 & 253.85  & 11.34 & 6.49  & 0.193 & 0.024  & 573 \\
$\textrm{d}_s$ & 2020-01-29 22:10 & 22.54 & 254.16 & 11.28  & 6.84  & 0.162 & 0.035  & 573 \\
$\textrm{e}_s$ & 2020-01-29 19:00 & -17.05 & 261.84 & 11.12  & 6.55  & 0.219 & 0.064  & 411 \\
$\textrm{f}_s$ & 2020-01-29 19:20 & -15.61 & 263.15 & 11.09  & 7.46  & 0.214 & 0.053  & 411 \\
$\textrm{g}_s$ & 2020-01-29 01:00 & 19.65 & 263.24 & 9.09  & 5.67  & 0.184 & 0.056  & 156 \\

\bottomrule
\end{tabular*}
\end{table*}

\section{Engineering Wake Modelling in LongSim}\label{models}

LongSim employs several analytically based models (with some empirical components) to represent various relevant wind turbine wake characteristics, such as velocity deficit, added turbulence, superposition and deflection. These models can be flexibly combined, allowing for numerous possible configurations. In this work, four distinct model combinations are selected for the sake of conciseness and clarity in the comparative analysis, as described below. While several modelling approaches are reviewed below for completeness, only the specific model combinations summarised in Table~\ref{tab1} are evaluated in this study.

\subsection{Wake Velocity Deficit} 

Three analytical wake-deficit formulations are reviewed here for context, although only a subset of them is used in the engineering model combinations introduced later in this section. (1) The model developed by Ainslie \cite{ainslie1988calculating} solves the RANS equations in two dimensions using cylindrical coordinates coupled with an eddy-viscosity model based on mixing-length theory. Ruisi and Bossanyi\cite{ruisi2019engineering} modified the eddy-viscosity term associated with momentum diffusivity to account for atmospheric stability by using the Monin-Obukhov length. (2) The model formulated by Bastankhah and Port{\'e}-Agel\cite{bastankhah2016experimental} is derived from mass and momentum conservation. It assumes self-similar Gaussian distributions aligned with the vertical and lateral axes relative to the plane normal to the flow direction, with a linear wake growth approximation, which can be tuned with the coefficients proposed by Niayifar \& Port{\'e}-Agel\cite{niayifar2016analytical}. Several extensions to this family of Gaussian wake models exist, including formulations that incorporate wind-direction veer\,\cite{abkar2018analytical}. This Gaussian model is reviewed here for completeness but is not used directly in the engineering model combinations. (3) The approach developed by Bastankhah et al\cite{bastankhah2022vortex} is based on the Biot–Savart law and the vorticity transport equation, representing the wake edge as a vortex sheet shed from the rotor disk circumference. This approach captures the time evolution of the wake's trajectory and modifies the Gaussian wake model into a curled wake shape. Mohammadi et al\cite{mohammadi2022curled} further refined the inflow conditions to incorporate veer, following the works of Abkar et al\cite{abkar2018analytical} and Narasimhan et al\cite{narasimhan2022effects}.

\subsection{Added Turbulence}

Four approaches to wake added turbulence are considered via empirical relationships between the ambient turbulence, rotor aerodynamic properties and turbine spacing. (1) Quarton and Ainslie \cite{quarton1990turbulence}, as modified by Hassan \cite{hassan1993wind}, modelled the added turbulence using empirical data relating wake turbulence to downstream distance, free-stream turbulence, and the rotor thrust coefficient. (2) Crespo and Hernandez\cite{crespo1996turbulence} parameterised the wake-added turbulence through empirical correlations, employing axial-induced velocity instead of the thrust coefficient, with validity limited to the far-wake region. (3) Zehtabiyan-Rezaie et al\cite{zehtabiyan2024wind} refined a coefficient and an exponent in Crespo and Hernandez's formulation using LES data. (4) Ishihara and Qian\cite{ishihara2018new} developed empirical relationships between ambient turbulence and rotor thrust coefficients based on LES data, including a double-Gaussian-shaped turbulence intensity distribution, assuming axial symmetry and self-similarity; in this work, this formulation is used only to prescribe the radial distribution of wake-added turbulence.

\subsection{Wake Superposition}

Two combinations of wake velocity deficits and wake-added turbulence were considered. (1) The dominant wake model \cite{gl2014theory} assumes the highest velocity deficit and added turbulence among all wakes affecting a turbine. (2) Bastankhah et al \cite{bastankhah2021analytical} integrated the velocity deficit superposition by cumulatively solving the flow-governing equations directly for a turbine immersed in upwind turbine wakes, where the turbulence superposition is calculated as the root sum of the squares of the ambient turbulence and the maximum wake-added turbulence components. Note that the cumulative wake model of Bastankhah et al \cite{bastankhah2021analytical} converges to the Gaussian wake model of Bastankhah and Port{\'e}-Agel \cite{bastankhah2016experimental} for a single-turbine scenario.

It is worth emphasising that wake superposition methods should be chosen in a manner that is physically consistent with the assumptions underlying the single-turbine wake formulation. In practice, engineering wake models are often combined with different superposition strategies without explicit consideration of their compatibility. Depending on the modelling assumptions, certain superposition approaches may be more appropriate than others, particularly under strong wake interaction. This consideration is especially relevant for practitioners applying engineering wake models in densely packed wind farms.

\subsection{Wake Deflection}

Three approaches for estimating the wake centreline's lateral deflection caused by yaw misalignment with the flow direction were considered. (1) The model developed by Bastankhah and Port{\'e}-Agel\cite{bastankhah2016experimental} estimates the lateral wake deflection for a given yaw angle by approximating the wake skew angle as a symmetric Gaussian distribution from the end of the potential core length based on Coleman's approximations\cite{coleman1945evaluation}. (2) The vortex sheet model from Bastankhah et al\cite{bastankhah2022vortex} predicts lateral deflection behind a yawed turbine using an equation that combines an approximately circular vortex sheet with the self-induced motion of a counter-rotating pair. (3) Gaukroger et al\cite{new} modelled wake deflection by estimating the yaw-induced lateral velocity as an asymmetric Gaussian distribution from the onset of the far wake, taking into account the cumulative effects of upwind turbine wakes, as in the methodology of Bastankhah et al\cite{bastankhah2021analytical}.

\subsection{Model Combinations}

Table \ref{tab:1} summarises the engineering model combinations for the four key components: velocity deficit, turbulence, superposition, and deflection. Each model is labelled (i, ii, iii and iv) to indicate a specific combination. Model i uses Ainslie's deficit \cite{ainslie1988calculating}, modified to include atmospheric stability effects \cite{ruisi2019engineering}. The velocity deficit and the added turbulence \cite{quarton1990turbulence,hassan1993wind}, with a double-Gaussian radial distribution \cite{ishihara2018new}, are combined using the dominant wake criterion \cite{gl2014theory}. Models i and ii share the same wake deflection approach \cite{bastankhah2016experimental}, whereas model ii calculates the velocity deficit via integrated wake superposition \cite{bastankhah2021analytical}. Models ii and iii use the same added deficit superposition method \cite{bastankhah2021analytical} without wake-added turbulence, while model iii considers a curled-wake deficit shape and deflection \cite{bastankhah2022vortex,mohammadi2022curled}. Finally, model iv uses the cumulative wake velocity deficit \cite{bastankhah2021analytical}, deflected by the yaw-induced lateral velocity \cite{new}, with a distinct added turbulence model \cite{crespo1996turbulence,zehtabiyan2024wind}, considering wake overlapping \cite{niayifar2016analytical} for partial wakes. The key governing equations and assumptions underlying the engineering wake models used in this study are provided in Appendix~A. 
\begin{table*}[!t]%
\centering %
\caption{Engineering model combinations evaluated.\label{tab1}}%
\begin{tabular*}{\textwidth}{@{\extracolsep\fill}lllll@{\extracolsep\fill}}
\toprule
\textbf{Model} & \textbf{Velocity Deficit}  & \textbf{Turbulence}  & \textbf{Superposition}  & \textbf{Deflection} \\
\midrule
i & Ainslie\cite{ainslie1988calculating} - Ruisi and Bossanyi \cite{ruisi2019engineering} & Quarton and Ainslie \cite{quarton1990turbulence}$^\dagger$ $^\ddagger$ & Dominant wake \cite{gl2014theory}  & Bastankhah \& Port{\'e}-Agel \cite{bastankhah2016experimental}   \\
ii & -  & -  & Bastankhah et al \cite{bastankhah2021analytical}  & Bastankhah and Port{\'e}-Agel \cite{bastankhah2016experimental}   \\
iii & Bastankhah et al\cite{bastankhah2022vortex} - Mohammadi et al \cite{mohammadi2022curled}  & -  & Bastankhah et al \cite{bastankhah2021analytical} & Bastankhah et al\cite{bastankhah2022vortex} - Mohammadi et al \cite{mohammadi2022curled}   \\
iv & -  & Crespo and Hernandez \cite{crespo1996turbulence} $^\S$  & Bastankhah et al \cite{bastankhah2021analytical}  & Gaukroger et al \cite{new}   \\
\bottomrule
\end{tabular*}
\begin{tablenotes}
\item[$^\dagger$] As modified by Hassan\cite{hassan1993wind}
\item [$^\ddagger$] TI distribution as defined by Ishihara \& Qian\cite{ishihara2018new}
\item[$^\S$] Updated with coefficients from Zehtabiyan-Rezaie et al \cite{zehtabiyan2024wind}

\end{tablenotes}
\label{tab:1}
\end{table*}

\section{Results}\label{validation}

This section examines the outcomes of the engineering wake-model combinations against Lillgrund field data, focusing on two main areas: wake measurements along the transect and the farm's power output. We assess how well the models capture velocity deficits and flow patterns downstream of the turbines. We also compare the predicted power output to assess how accurately the models reproduce the observed relative power variations across the array for the selected cases. The following analysis evaluates the accuracy of each model and highlights discrepancies with field data for normal operation and wake steering.

\subsection{Transect Velocity Deficit}

The transect spans approximately eight rotor diameters in the \(\hat{y}\) direction, as illustrated in Figure \ref{fig_layout}b. The wind directions in the selected cases result in wind farm layouts where the transect lies within the wake of the third column, with the nearest upstream turbines located approximately one to three rotor diameters upstream. Therefore, we examine the wake velocity deficit during the transition from the near-wake region to the far-wake region, considering the range of inflow and operating conditions represented in the selected cases. 
\begin{figure*}[htbp]
\centerline{\includegraphics[width=0.9\textwidth]{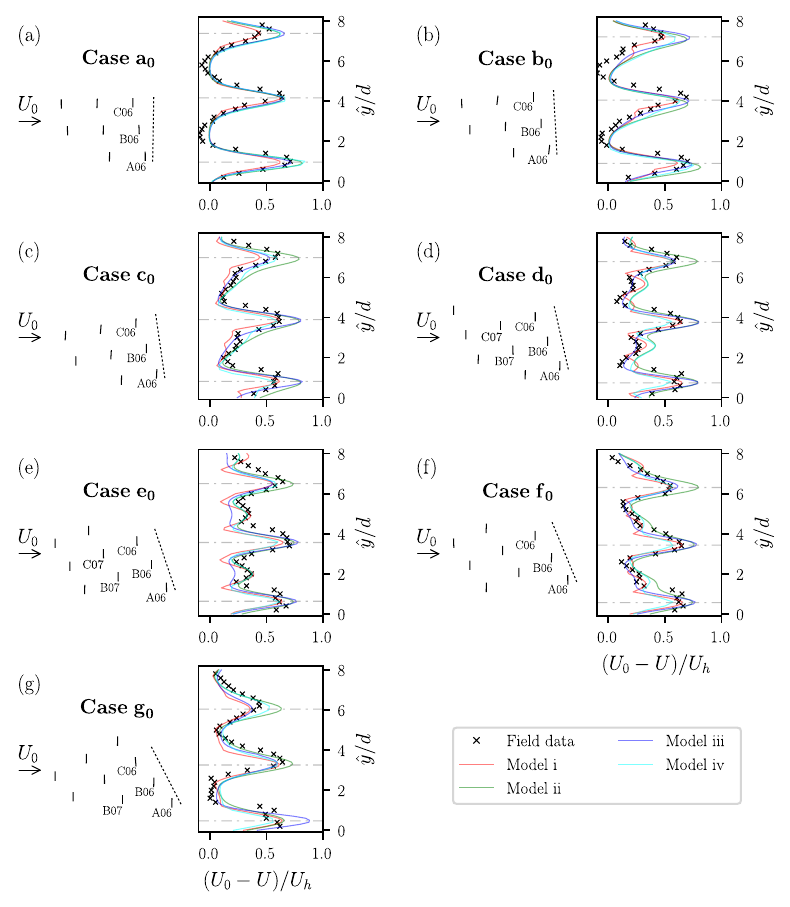}}
\caption{Streamwise normalised velocity deficit profiles at the transects (a) to (g) for the corresponding baseline cases \( \textrm{a}_0 \) to \( \textrm{g}_0 \). The case layout is shown on the left, where the solid lines indicate the rotor footprints of turbines positioned directly upstream of the transect, represented by the dotted line. The horizontal dot-dashed lines denote the streamwise projection of the rotor centres of the nearest upstream turbines onto the transect. \label{fig_transect_profile_0}}
\end{figure*}

\subsubsection{Velocity Deficit in the Baseline Cases}

Figure~\ref{fig_transect_profile_0} presents the transect velocity deficit for the baseline cases. Each subplot includes a schematic of the upstream turbine layout relative to the transect, where turbines are marked with solid lines and a dotted line indicates the transect location. For consistency, the incoming wind direction \( U_0 \) is reoriented from left to right in all cases. Any apparent yaw offsets shown in the schematics are based on the normal operation of the wind turbine controllers (with no wake steering), and they reflect the nacelle orientation relative to the prescribed wind direction in each selected scenario. Overall, the analytical models capture the main deficit patterns observed in the field measurements across the sampled wind directions. For example, in case~\(\textrm{a}_0\), the models reproduce full-wake conditions, with peak velocity deficits observed directly downstream of the third-column turbines A06, B06 and C06 and reduced deficits in the regions between turbines. As the wind direction varies systematically across the other scenarios, the models generally capture the influence of second-column wakes on the transect profiles, particularly between turbines A06 and B06 and between B06 and C06. Notably, in cases~\(\textrm{d}_0\) and~\(\textrm{e}_0\), local peak deficits appear in these regions, aligned with the downstream positions of turbines B07 and C07.

The modelled velocity deficit varies according to the assumptions embedded in each engineering model combination. Models~ii and iii tend to overestimate the maximum deficit relative to models~i and iv, reflecting differences in their treatment of wake superposition and deficit formulation. Distinctively, the transect profiles from model~i exhibit sharper transitions compared to the other models, likely due to the dominant wake criterion used for superposition. Other discrepancies with the field data are case-specific. For example, model~iii estimates the local peak deficit downstream of turbine C07 more accurately than the other models in case~\(\textrm{d}_0\) but underestimates the same feature in case~\(\textrm{e}_0\). This contrast underscores the scenario-dependent behaviour of the models and the sensitivity of their parameters to varying flow conditions.

These discrepancies are particularly evident in the near-wake region, where the transect lies relatively close to the nearest upstream turbine. This configuration presents a known challenge for engineering wake models, which tend to underperform in this region due to the presence of strong velocity gradients and incomplete wake development. The Gaussian-shaped deficit predicted by most models provides a reasonable approximation of the field data, except at the location of the downstream turbine A06, which also lies within the near-wake region in most cases. In this context, a super-Gaussian formulation \cite{blondel2020alternative} may yield a better fit. 

In addition to near-wake limitations, other flow features not explicitly resolved by the models may contribute to the observed discrepancies. The current framework does not account for inflow velocity non-uniformity or local flow acceleration (or speed-up), both of which are known to influence wake evolution. Under full-wake conditions in Figure~\ref{fig_transect_profile_0}a, variations in the inter-turbine regions are presumably due to inflow heterogeneity and/or flow speed-up between adjacent turbine rows. A similar effect is visible in the nearly aligned case of Figure~\ref{fig_transect_profile_0}g, where the upstream turbine B07 is non-operational, and a speed-up occurs between turbines A06 and B06. Flow acceleration is also evident in Figure~\ref{fig_transect_profile_0}b, where a lateral displacement between the modelled and measured deficit peaks may be attributed to directional inflow variability, consistent with the relatively large average absolute yaw angle~\(\Gamma\) in case~\(\textrm{b}_0\).

\subsubsection{Velocity Deficit in the Steering Cases}

\begin{figure*}[htbp]
\centerline{\includegraphics[width=0.9\textwidth]{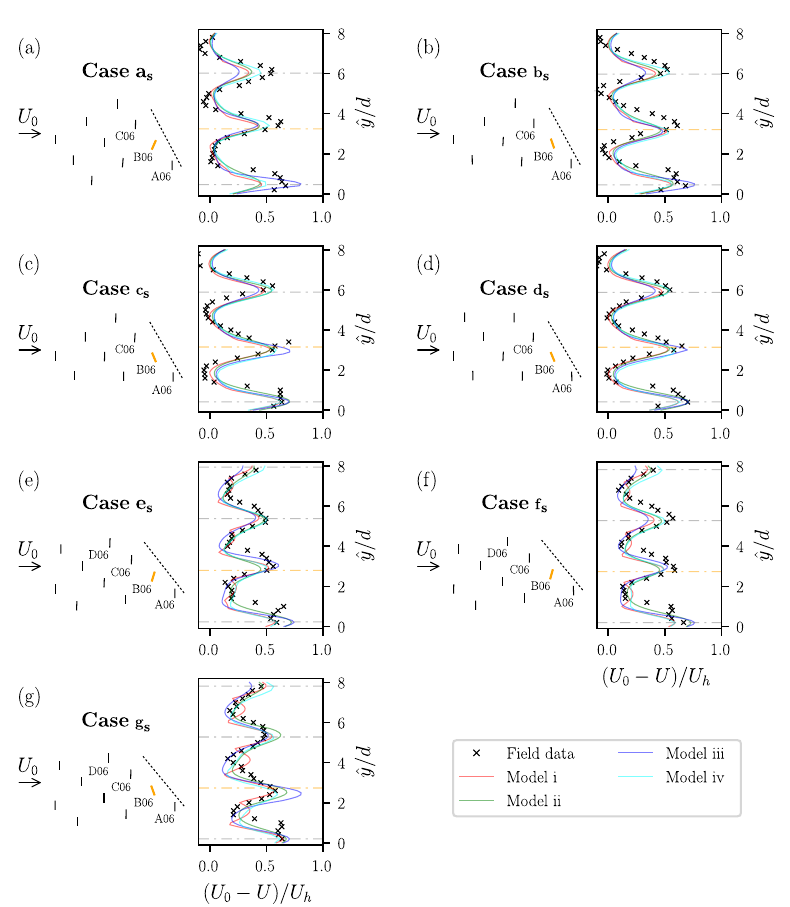}}
\caption{Streamwise normalised velocity deficit profiles at the transects (a) to (g) for the corresponding steering cases \( \textrm{a}_s \) to \( \textrm{g}_s \). The case layout is shown on the left, where the solid lines indicate the rotor footprints of turbines positioned directly upstream of the transect, represented by the dotted line. The horizontal dot-dashed lines denote the streamwise projection of the rotor centres of the nearest upstream turbines onto the transect. \label{fig_transect_profile_s}}
\end{figure*}

Figure~\ref{fig_transect_profile_s} presents the transect velocity deficits for the wake steering cases. Each subplot also shows the rotor layout and transect location, with the intentionally yawed turbine highlighted in a different colour for clarity. The overall deficit profiles are broadly consistent with those in the baseline cases. Configurations with greater overlap between turbine rows are shown in Figure~\ref{fig_transect_profile_s}a--d, while more staggered layouts are depicted in Figure~\ref{fig_transect_profile_s}e--g. The analytical models generally reproduce the yaw-induced wake deflection downstream of turbine~B06, located approximately two rotor diameters upstream of the transect. Across cases, the observed variation in the wake centre position reflects the sensitivity of deflection predictions \cite{bastankhah2016experimental,new} to the assumed wake width growth in each formulation. The wake shifts towards the positive transect direction for clockwise yaw angles, as seen in cases~$\textrm{a}_s$, $\textrm{e}_s$ and~$\textrm{f}_s$, and towards the negative direction for anticlockwise yaw angles in cases~$\textrm{c}_s$, $\textrm{d}_s$, and~$\textrm{g}_s$. Here, clockwise and anticlockwise yaw correspond to negative and positive
$\gamma_{\mathrm{B06}}$, respectively (see Section~2.4). An exception is case~$\textrm{b}_s$, where the wake appears laterally displaced in the opposite direction, which may reflect spatial inhomogeneity in the inflow direction. Other discrepancies between the modelled and measured profiles are attributed to the sensitivity of model parameters to flow conditions, as well as to inherent limitations of the modelling framework. For instance, models~ii and~iii overestimate the maximum deficit in case~$\textrm{g}_s$, while the minimum deficits for cases~$\textrm{a}_s$--$\textrm{d}_s$ do not capture the flow acceleration observed in the field data.

Beyond the influence of yaw angle alone, upstream wakes can also alter the deflection of the wake from turbine~B06, depending on the relative position of upstream turbines \cite{zong2021experimental}. In the presence of lateral offsets between turbines, the superposition of upstream wakes can distort the local velocity field into which the yawed turbine emits its wake, thereby biasing the resulting deflection. In aligned configurations, the deflection induced by a yawed turbine typically steers the entire cumulative wake in the same direction. However, when turbines are laterally offset, the interaction between the yaw-induced deflection and the background velocity field, shaped by upstream wakes, can either amplify or suppress the apparent wake shift, depending on the relative positions of the turbines and their yaw direction. Even in the absence of yaw, the combined influence of multiple laterally offset wakes can generate a net lateral drift of the overall wake field, depending on the layout geometry and inflow alignment. This upstream influence may help explain the asymmetry observed between clockwise and anticlockwise yaw cases. For example, in case~$\textrm{a}_s$, the projection of upstream turbines lies on the same side as the expected deflection, effectively reinforcing the shift of the wake in the positive transect direction. In contrast, in cases~$\textrm{c}_s$ and~$\textrm{d}_s$, the upstream wakes appear to bias the flow in the opposite direction, counteracting the yaw-induced deflection and resulting in a reduced lateral shift. Another illustrative example arises from cases~$\textrm{e}_s$ and~$\textrm{g}_s$, which share similar layout configurations and yaw magnitudes of opposite signs. Though the wake asymmetries are less pronounced than in cases with greater wake overlap, the upstream turbines still appear to modulate the effective deflection direction: case~$\textrm{g}_s$, the upstream wakes reinforce the yaw-induced shift, while in case~$\textrm{e}_s$, they appear to oppose it.

\begin{figure*}[htbp]
\centerline{\includegraphics[width=0.8\textwidth]{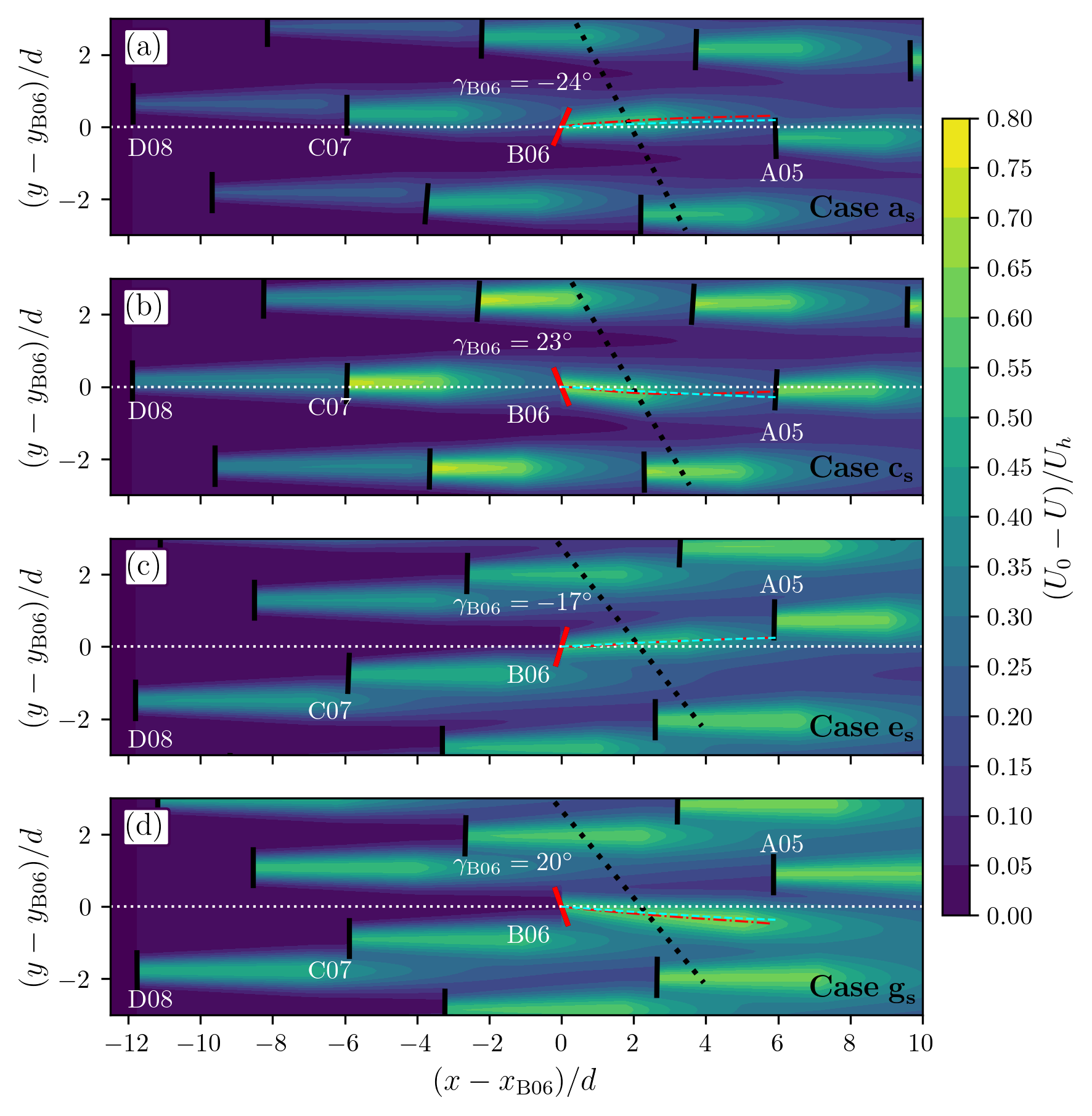}}
\caption{Normalised velocity deficit contours from model~iv for steering cases~$\textrm{a}_s$~(a), $\textrm{c}_s$~(b), $\textrm{e}_s$~(c) and $\textrm{g}_s$~(d). Rotor footprints (solid lines) and the transect (black dotted line) are shown relative to the position of turbine~B06, whose rotor is highlighted in red to indicate intentional yaw misalignment. A white dotted line marks the centreline of B06. Additionally, the red dash-dot line traces the downstream wake centre where the maximum deficit occurs under the influence of B06, while the cyan dashed line indicates the estimated wake centre in the absence of upstream turbine effects. The yaw angle of B06, \( \gamma_{\mathrm{B06}} \), is shown in each subplot and rounded to the nearest integer. \label{fig:model_iv_contours}}
\end{figure*}

To further illustrate these trends, Figure~\ref{fig:model_iv_contours} presents normalised velocity deficit contours from model~iv for cases~$\textrm{a}_s$, $\textrm{c}_s$, $\textrm{e}_s$ and~$\textrm{g}_s$. The rotor footprints (solid lines) and transect (black dotted line) are shown relative to the position of turbine~B06, with a white dotted line indicating its projected centreline in the inflow direction. A red dash-dot line traces the downstream path of maximum velocity deficit, representing the actual wake centre under the influence of yaw and upstream interactions, while a cyan dashed line denotes the estimated wake centre in the absence of upstream turbine effects. In case~$\textrm{a}_s$, the wakes from upstream turbines D08 and C07 deepen the deficit on the same side as the expected deflection, lowering the local velocity and thereby reinforcing the yaw-induced wake shift. In contrast, in case~$\textrm{c}_s$, the cumulative deficit on the deflection side is weaker, resulting in a more limited lateral displacement. A similar, though less pronounced, pattern appears in cases~$\textrm{e}_s$ and~$\textrm{g}_s$, where the greater spanwise offset of upstream turbines diminishes their influence but does not fully eliminate the asymmetry in wake deflection. We note that this upstream influence is a modelled feature and is not claimed to be directly observable in the field measurements.

\subsubsection{Error Analysis of Velocity Deficit Predictions}

\begin{figure*}[htbp]
\centerline{\includegraphics[width=0.45\textwidth]{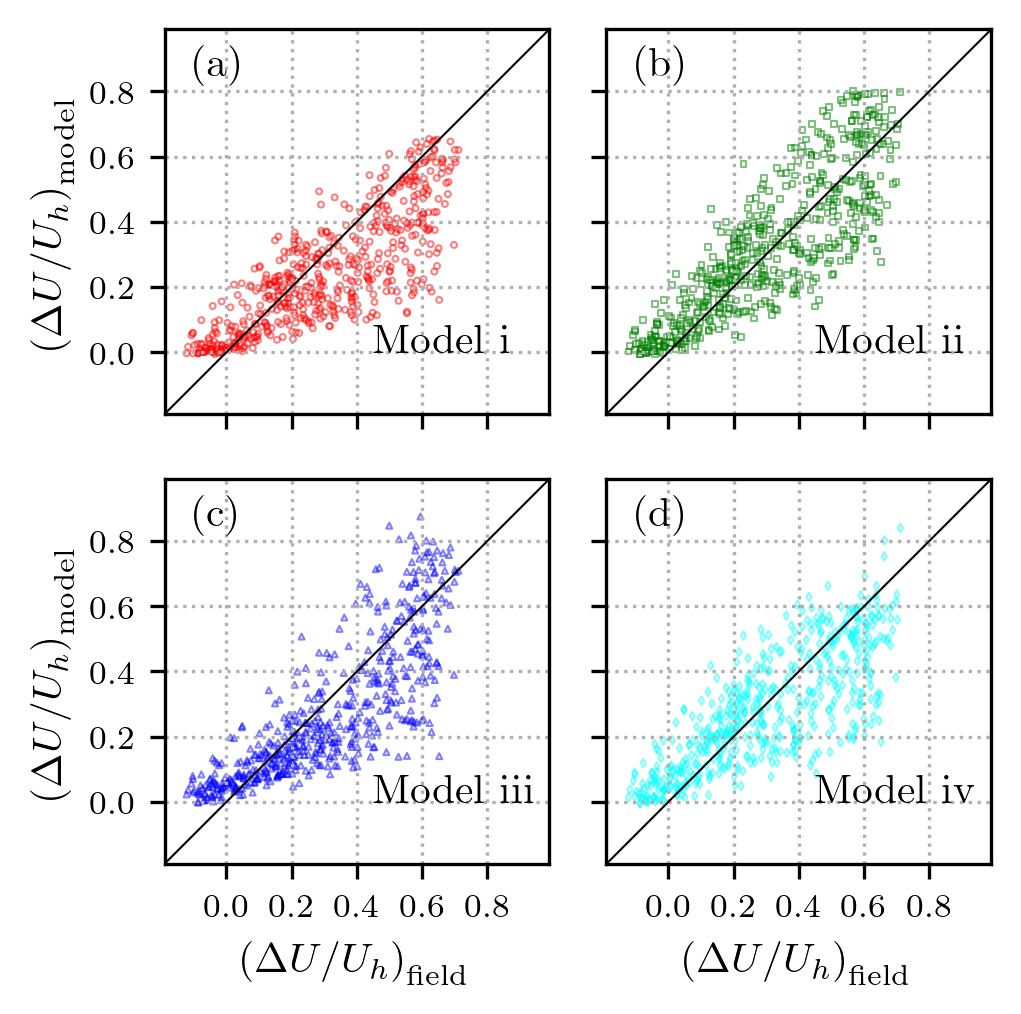}}
\caption{Transect streamwise normalised velocity deficit relative to field data for all cases: (a) model i, (b) model ii, (c) model iii and (d) model iv. \label{fig_t_scatter}}
\end{figure*}

The normalised velocity deficit at the transect for each model, derived from all baseline cases and steering scenarios, is collected to evaluate model performance relative to the field data, as shown in Figure~\ref{fig_t_scatter}. The deficit is defined as \(\Delta U = U_0 - U\). As previously noted, the models exhibit limitations in capturing localised speed-ups (\(U > U_0\)) visible in the lower-left region, likely resulting from flow channelling between turbine columns. The modelled data become increasingly scattered at higher deficit values, reflecting larger inaccuracies. These errors are partly attributed to known modelling challenges in the near-wake region downstream of turbine A06, where engineering wake models often fail to capture complexities associated with the near-wake region. Overall, models~i and~iv tend to underestimate the maximum deficits (e.g., see case $\textrm{e}_0$ in Figure~\ref{fig_transect_profile_0}), while models~ii and~iii either underestimate or overestimate them (e.g., see case $\textrm{a}_s$ in Figure~\ref{fig_transect_profile_s}), with model~ii showing relatively better accuracy at peak deficits. Additional sources of error may stem from assumptions such as inflow homogeneity, which limit the ability to represent complex flow features present in the field.

\begin{figure*}[htbp]
\centerline{\includegraphics[width=0.9\textwidth]{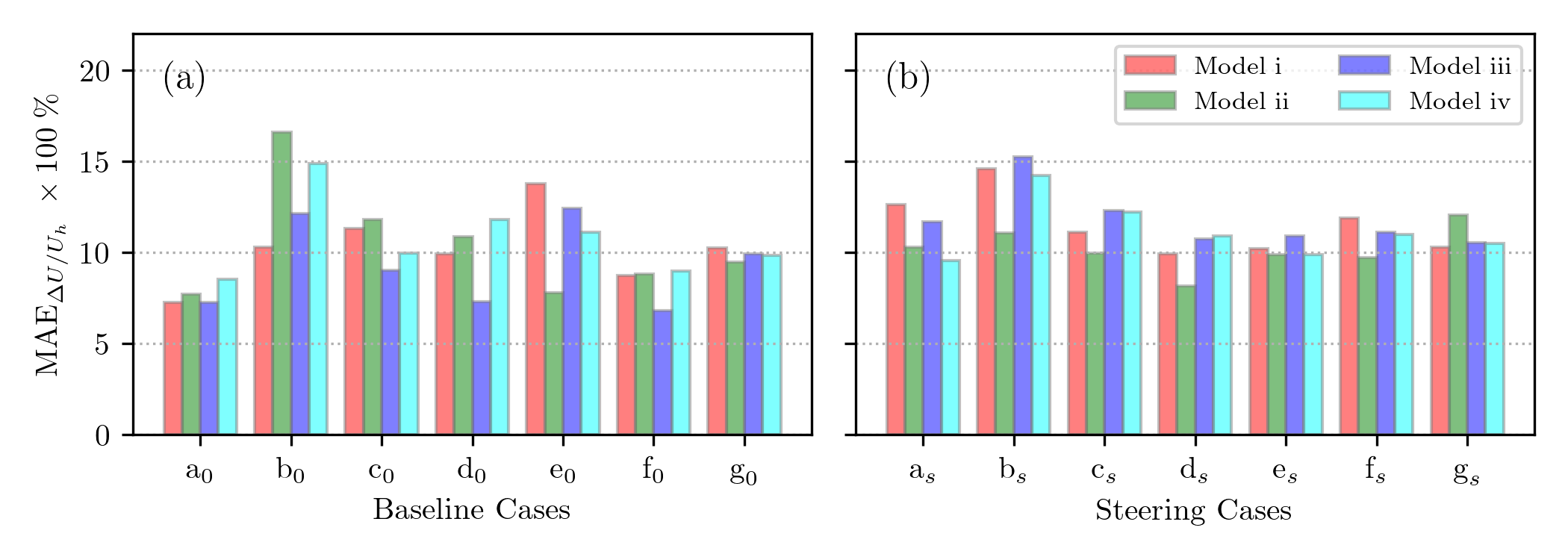}}
\caption{MAE of the normalised velocity deficit at the transect for the (a) baseline cases and (b) steering cases. The legend in (b) applies to both subfigures. \label{fig_mae_u}}
\end{figure*}

To complement this analysis, the accuracy of transect deficit modelling on a per-case basis is quantified using the mean absolute error (MAE), which represents the average magnitude of absolute differences between modelled and observed normalised values, as shown in Figure~\ref{fig_mae_u}. The accuracy is both case- and model-dependent, with no apparent performance advantage observed under either aligned or non-aligned farm layouts or between baseline and steering scenarios. Overall, most scenarios yield an error of approximately 10\%, with a minimum around 7\% in case~\(\textrm{a}_0\) across all four models. The highest MAE values, close to 15\%, are found in cases such as~\(\textrm{b}_0\) and~\(\textrm{b}_s\), where the inflow is likely less homogeneous, and discrepancies may arise from local flow variations and uncertainty in the estimated wind direction across the farm.

\subsection{Power Predictions}

The modelled power is compared with field data to evaluate the performance of the engineering models from a yield assessment perspective. Turbine power is indirectly estimated using the rotor-averaged streamwise velocity derived from the analytical models in the location of the rotor (in the absence of turbine), combined with interpolated power curves corresponding to the specified yaw and pitch angles, as shown in Figure~\ref{fig_turbine}c. Power production for all turbines are normalised by \( P_{\mathrm{D08}} \), defined as the power of the most upstream turbine, D08. To evaluate wake-induced losses independently from absolute power levels,  we normalise each turbine´s power using the ratio $(P/P_{\mathrm{D08}})_{\mathrm{field}}$ for the measurements and $(P/P_{\mathrm{D08}})_{\mathrm{model}}$ for the model outputs. This approach enables a consistent comparison of relative power reductions across the array while mitigating the impact of inflow discrepancies between the model and measurements. The results are assessed both qualitatively and quantitatively for the baseline and steering scenarios. Accuracy is evaluated using the MAE of the normalised power and the overall wind farm power prediction error of the models.

\subsubsection{Power Predictions in the Baseline Cases}

\begin{figure*}[htbp]
\centerline{\includegraphics[width=0.8\textwidth]{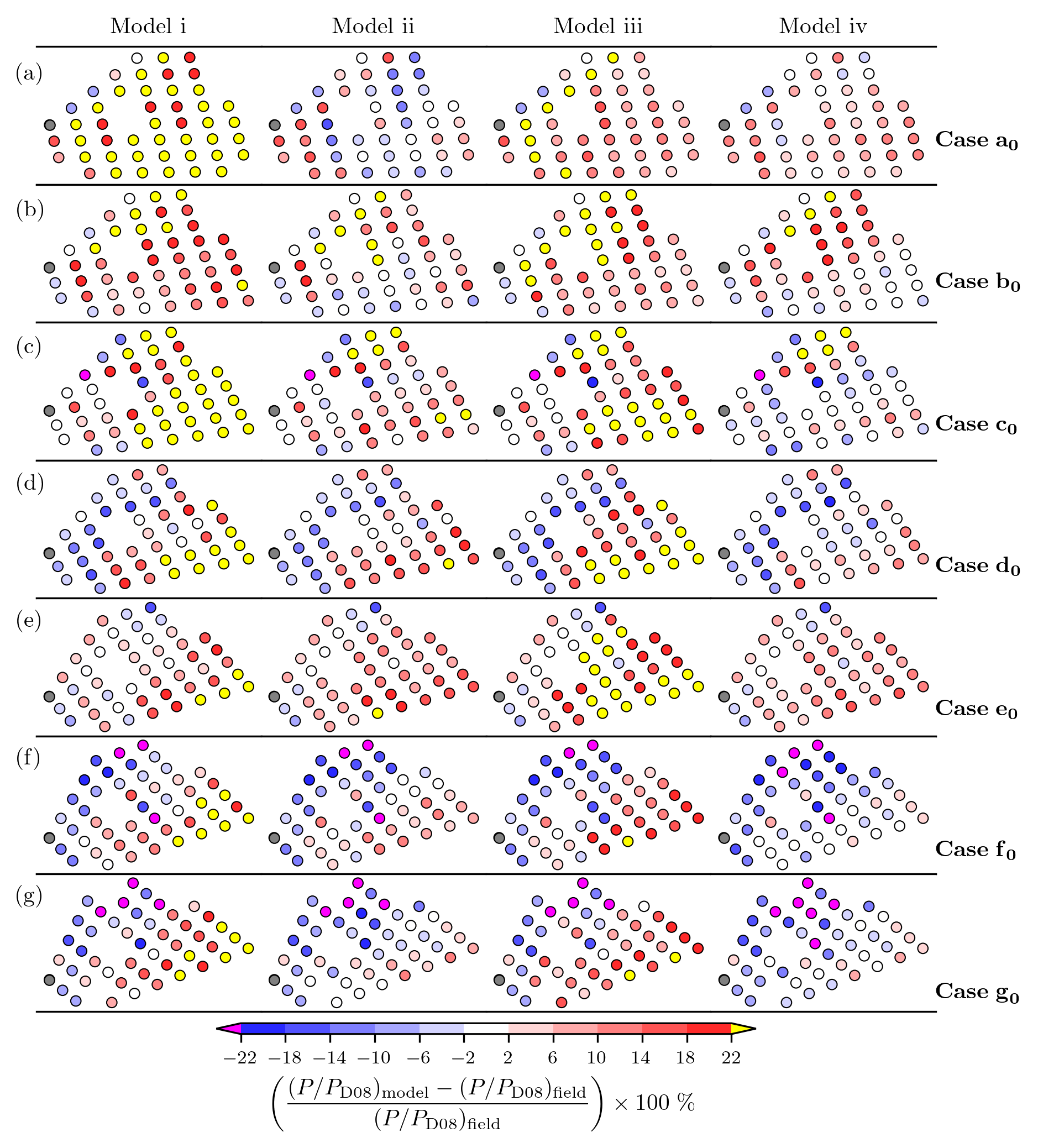}}
\caption{Wind farm layout and normalised power difference contours relative to the upstream turbine D08, shown in panels~(a) to~(g) for the corresponding baseline cases~$\textrm{a}_0$ to~$\textrm{g}_0$. The reference turbine D08 is shaded in grey. \label{fig_p_contour_0}}
\end{figure*}

Figure~\ref{fig_p_contour_0} presents the wind farm layouts for the baseline cases, rotated to align with a \(270^\circ\) inflow direction, consistent with the schematics in Figure~\ref{fig_transect_profile_0}. The contours show the power difference between the modelled and field data, normalised by the power of turbine D08. Although the contours vary across models and cases, deviations in the upstream turbines on the left side of each layout reveal evidence of non-uniform inflow, likely influenced by spatial heterogeneity and/or blockage effects. These features are not represented in the evaluated analytical models, which assume uniform, undisturbed inflow. For instance, the underestimation of upstream power in case~\(\textrm{a}_0\) suggests an underprediction of incoming wind speed, consistent with the minimal velocity deficits observed in Figure~\ref{fig_transect_profile_0}a. Slight discrepancies in the upstream contour pattern for model~iv also reflect sensitivity to the discretised wind farm grid, which affects the rotor-averaged wind speed estimation.

Further downstream, the power discrepancies tend to grow, highlighting how wake interactions accumulate through the farm. This is consistent with the compounding effect of wake deficits on power production, a feature that engineering models struggle to reproduce under spatially varying conditions. The appearance of localised outliers in the upper section of the farm, particularly where inflow measurements were taken from the opposite boundary, further underscores the role of inflow asymmetries and directionality.

\begin{figure*}[htbp]
\centerline{\includegraphics[width=0.8\textwidth]{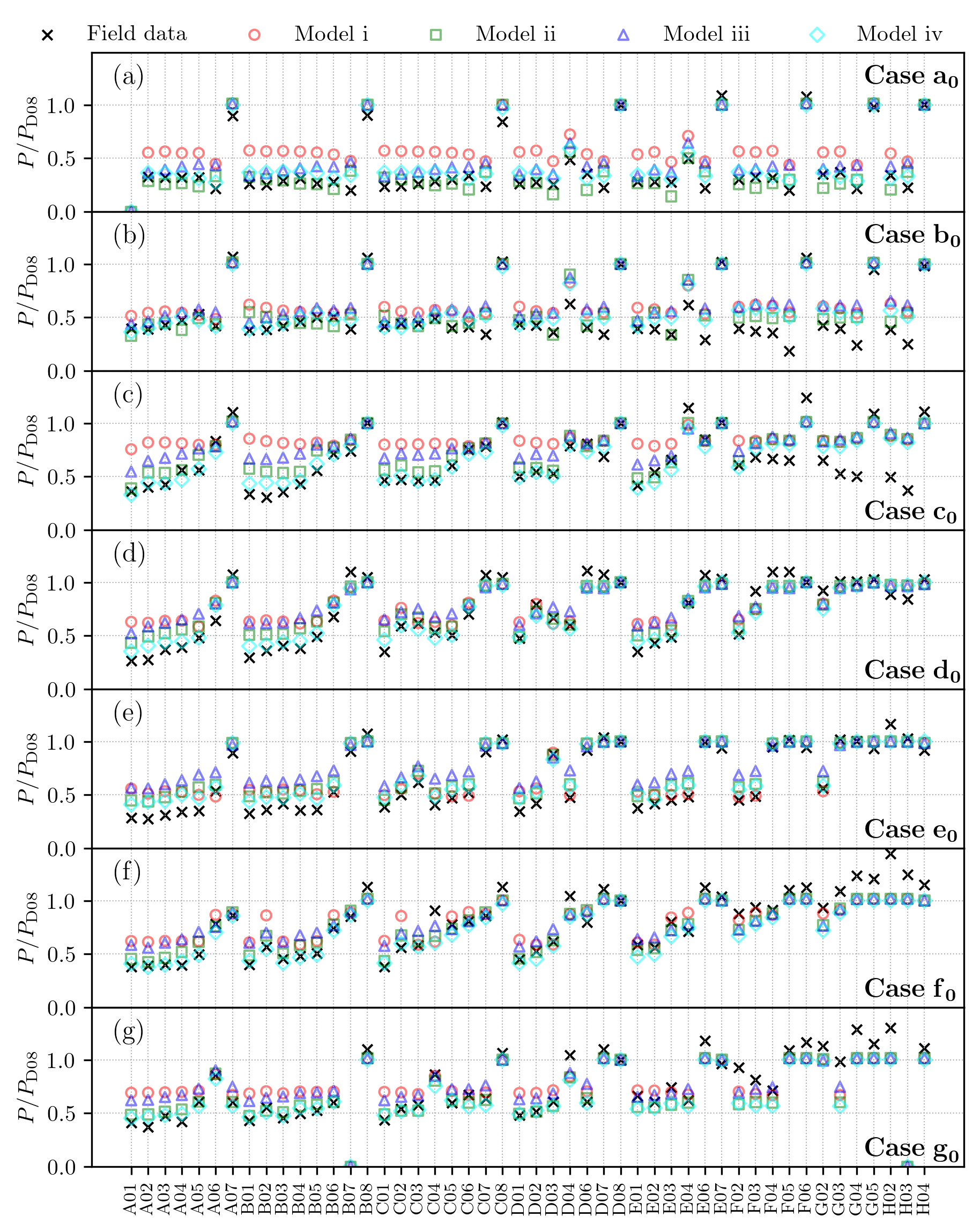}}
\caption{Relative power compared to the power of the upstream turbine, D08, (a) to (g) for the corresponding baseline cases \( \textrm{a}_0 \) to \( \textrm{g}_0 \).\label{fig_p_scatter_0}}
\end{figure*}

Figure~\ref{fig_p_scatter_0} provides a quantitative comparison of the modelled and measured turbine power for the baseline cases, expressed relative to \( P_{\mathrm{D08}} \). The models generally capture the turbine-level power trends as a function of wake exposure, with models~ii and~iv showing improved agreement for turbines situated deeper within the array. Notably, upstream power ratios deviating from unity point to persistent inflow heterogeneity, particularly affecting the northernmost turbines, shown on the right side of the figure. These patterns suggest that while engineering wake models can reproduce primary wake behaviour, their simplified assumptions limit accuracy in regions where flow acceleration, shear or recovery deviate from the idealised formulations. 

\subsubsection{Power Predictions in the Steering Cases}

\begin{figure*}[htbp]
\centerline{\includegraphics[width=0.8\textwidth]{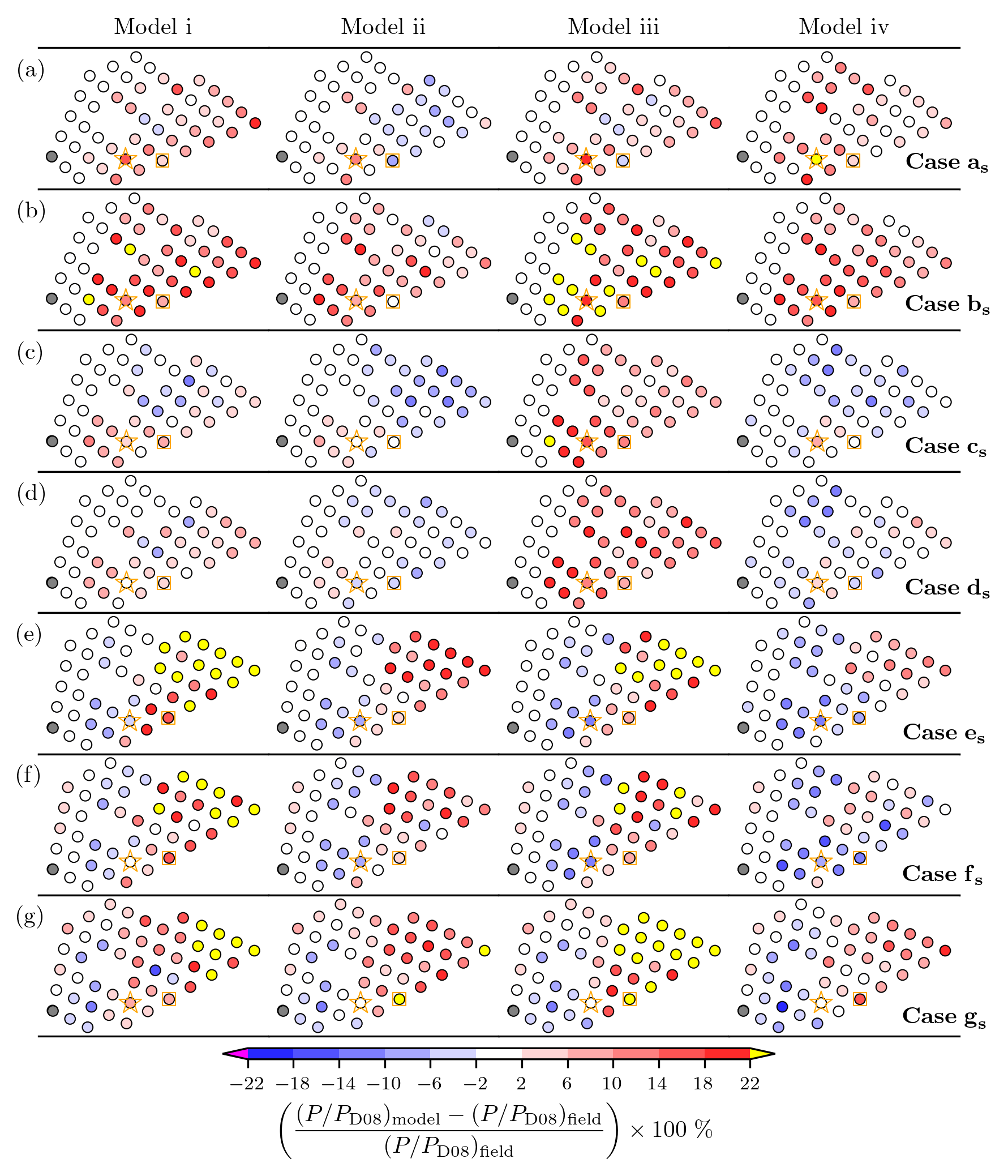}}
\caption{Wind farm layout and normalised power difference contours relative to the upstream turbine D08, shown in panels~(a) to~(g) for the corresponding steering cases~$\textrm{a}_s$ to~$\textrm{g}_s$. The reference turbine D08 is shaded in grey. The star symbol represents the steered turbine B06, and the square symbol represents its downstream turbine A05.\label{fig_p_contour_s}}
\end{figure*}

\begin{figure*}[htbp]
\centerline{\includegraphics[width=0.8\textwidth]{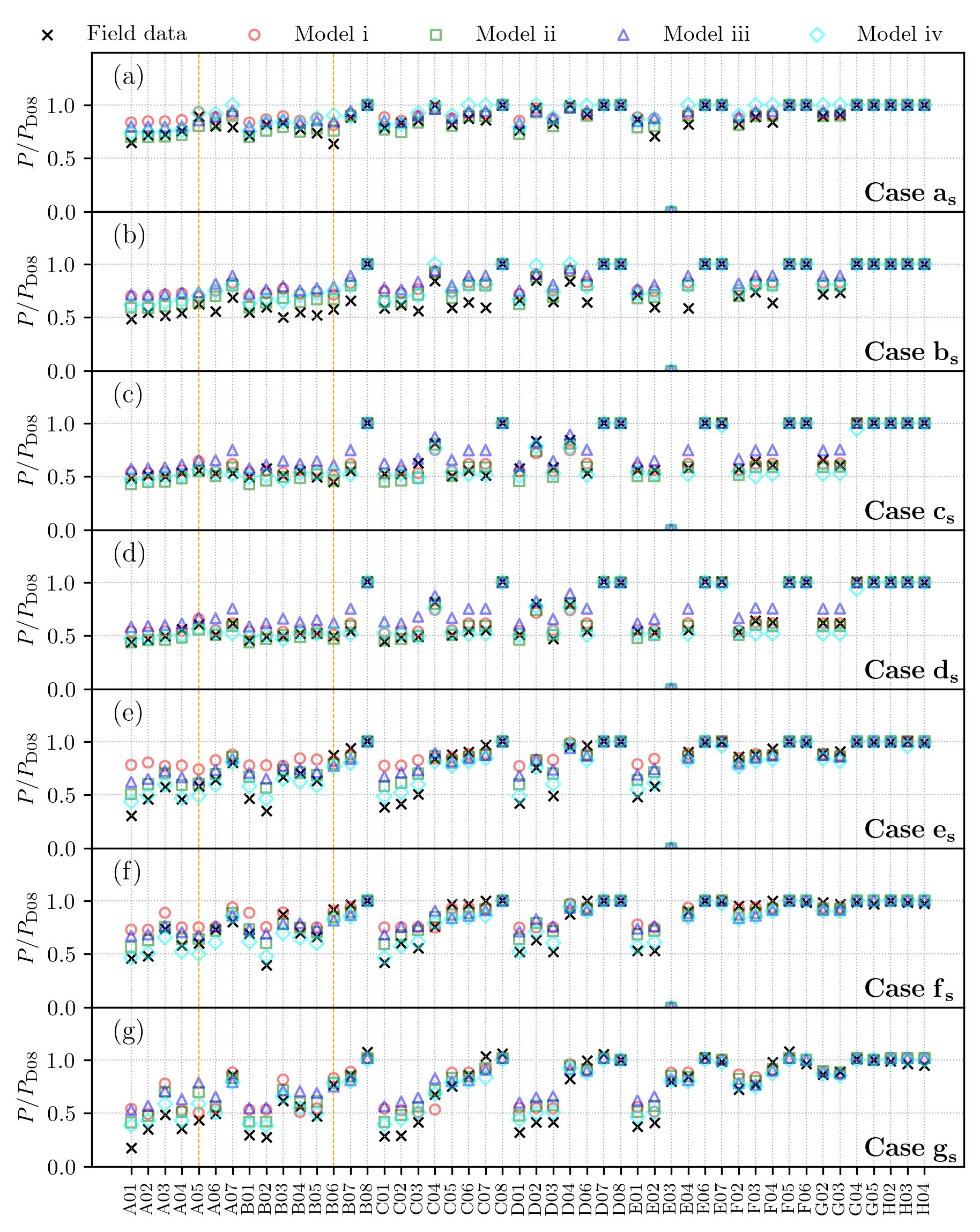}}
\caption{Relative power compared to the power of the upstream turbine, D08, for the steering cases (a) to (g). The vertical dashed lines represents the steered turbine B06, and its downstream turbine A05.\label{fig_p_scatter_s}}
\end{figure*}

The qualitative and quantitative power distributions for the steering cases are shown in Figures~\ref{fig_p_contour_s} and~\ref{fig_p_scatter_s}, respectively. Across the steering cases (except~\(\textrm{g}_s\)), the upstream turbines exhibit more uniform power levels than in several of the baseline cases. One contributing factor may be that inflow velocities during the steering cases are generally higher and closer to rated conditions (see Tables~\ref{tab2} and~\ref{tab3}). Under such conditions, many turbines operate near or at their rated power, where the power output becomes less sensitive to spatial variations in inflow velocity due to the saturation effect of the power curve. Consequently, moderate spatial heterogeneity in inflow is less clearly expressed in the power response. We emphasise that the baseline and steering datasets differ systematically in operating regime and atmospheric conditions (Section~\ref{lillgrund}), so the contrasts discussed here are descriptive rather than a controlled attribution to wake steering alone. Further downstream, however, the discrepancy between the modelled and measured power increases, consistent with the compounding effect of wake interactions, which encompass both cumulative downstream deficits and localised flow modifications induced by the yawed turbine. This tendency is captured to varying degrees by the models, with models~ii and~iv aligning more closely with the field data in reproducing the overall distribution of power across the array, as shown in Figure~\ref{fig_p_scatter_s}. In particular, case~$g_s$ exhibits a more spatially coherent positive power prediction error in Figure~\ref{fig_p_contour_s}. The corresponding velocity-deficit transects (Figure~\ref{fig_transect_profile_s}) indicate a slower downstream decay of the measured wake deficits than predicted by the models at several interaction locations, particularly after successive wake interactions. This behaviour is consistent with limitations in the analytical wake-interaction and recovery parameterisations and cannot be unambiguously attributed to farm-scale confinement effects based on the available evidence. In compact arrays such as Lillgrund, where wake interaction occurs before classical far-wake conditions are reached, the assumption of linear or weakly linear wake expansion, commonly adopted in engineering models, may further limit their ability to represent wake growth and recovery under repeated wake interactions.

A more localised view of the yaw steering effect can be seen in the power response of the steered turbine B06 and its downstream neighbour A05. The deficit examples from model~iv in Figure~\ref{fig:model_iv_contours} illustrate how wake deflection from B06 influences the downstream flow, with a visible lateral displacement of the wake that affects the power levels of both turbines. These effects are qualitatively represented in Figure~\ref{fig_p_contour_s}, where B06 and A05 are marked with star and square symbols, respectively. Throughout all steering cases, inaccuracies in the predicted power at B06 are compounded by discrepancies inherited from upstream turbines, which in turn cascade into the estimation of A05's power output. This behaviour is further illustrated in the quantitative results shown in Figure~\ref{fig_p_scatter_s}, where B06 and A05 are indicated by vertical dashed lines, allowing for direct comparison of their responses across models and cases. Most models capture the general redistribution of power between the two turbines, with improved agreement at A05 suggesting that the gross features of wake deflection are represented. However, larger discrepancies may arise from limitations in yaw-angle-dependent power characterisation and accumulated upstream wake errors. Across all steering cases, no model consistently outperforms the others at predicting the power at B06 or A05, indicating that the treatment of local wake dynamics under variable operating conditions remains challenging. This is especially the case for the models tested in this work as they are all steady-state engineering models. A better agreement could potentially be achieved using dynamic wake models \cite{becker2022revised,starke2024dynamic,becker2022ensemble,larsen2008wake}; however, these models require more accurate inflow condition data, which is not available to us.

\subsubsection{Error Analysis of Power Predictions}

\begin{figure*}[htbp]
\centerline{\includegraphics[width=0.9\textwidth]{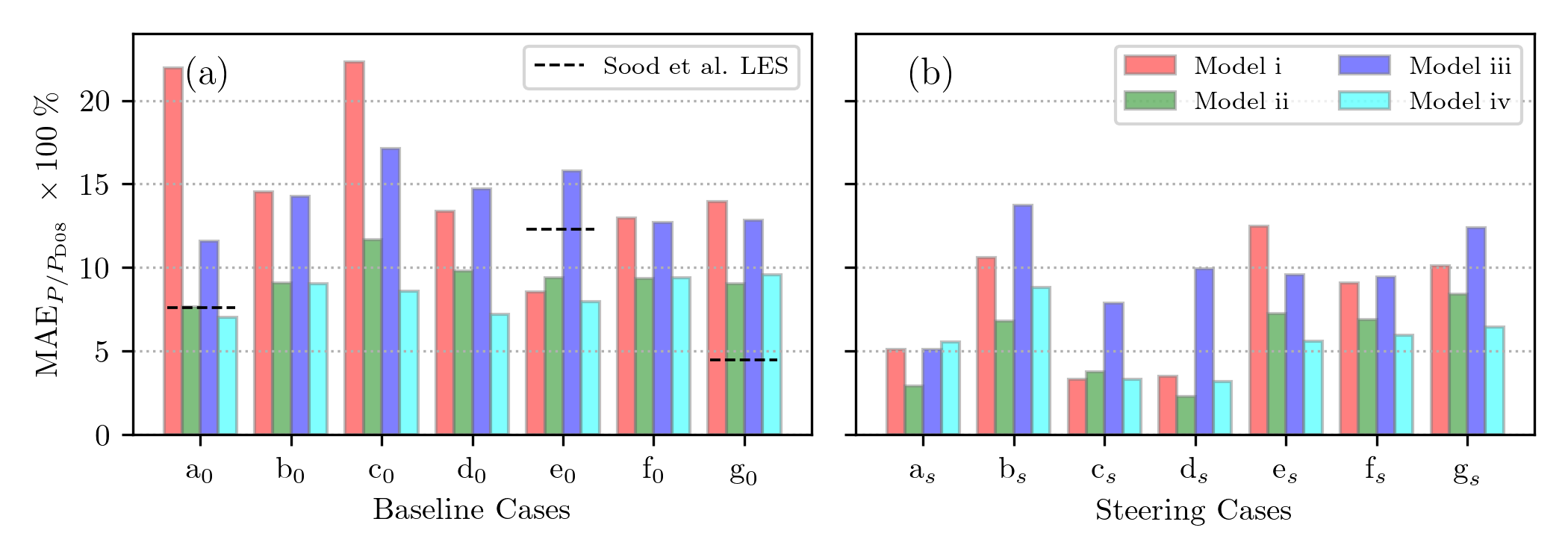}}
\caption{MAE of the normalised power for the (a) baseline cases and (b) steering cases. Dashed lines represent three LES cases from Sood et al \cite{sood2022comparison}. The legend in (b) applies to both subfigures. \label{fig_mae_p}}
\end{figure*}

Figure~\ref{fig_mae_p} presents the MAE of the normalised power relative to a reference value, highlighting differences in model performance across baseline and steering scenarios. In the baseline cases, model accuracy is generally comparable across configurations with varying wake influences (i.e., those resembling full- or partial-wake conditions). In the steering cases, MAE values tend to be lower than in many of the baseline cases; this trend may partly reflect operation closer to rated wind speeds, where turbine power becomes less sensitive to moderate inflow variability.

When comparing individual model performance, models~i and~iii exhibit greater variability, with MAE values occasionally dropping to around 5\% in cases such as~$\textrm{a}_s$, but often exceeding 10\%. In contrast, models~ii and~iv consistently yield lower errors, ranging between 3\% and 10\%, indicating better predictive reliability across both baseline and steering cases.

For broader context, three of the five LES cases simulated by Sood et al~\cite{sood2022comparison} corresponded to wind directions similar to those analysed here. Specifically, LES cases CNK8, PDK2 and CNK4, with inflow directions of \(222^\circ\), \(243^\circ\) and \(251^\circ\), respectively, approximate those of cases~$\textrm{a}_0$, $\textrm{e}_0$ and $\textrm{g}_0$. For these LES cases, the reported hub-height inflow wind speeds and turbulence intensities are $U_h=\{10.2,\,8.5,\,10.8\}\,\mathrm{m\,s^{-1}}$ and
$\mathrm{TI}=\{5.73,\,6.27,\,5.59\}\%$ for CNK8, PDK2 and CNK4, respectively, which are comparable in magnitude to the inflow conditions of the corresponding baseline field cases (cases~$a_0$, $e_0$ and~$g_0$ in Table~\ref{tab2}) in terms of hub-height wind speed and turbulence intensity. The LES simulations do not correspond to the same time periods as the field data analysed here and do not reproduce the full atmospheric state, but are included to provide contextual reference for model-data discrepancies under similar inflow regimes. As shown in Figure~\ref{fig_mae_p}, the associated MAE values from the LES results range from 4.5\% to 12.3\%, which are comparable to the errors obtained from the analytical models in this study.

\begin{figure*}[htbp]
\centerline{\includegraphics[width=0.9\textwidth]{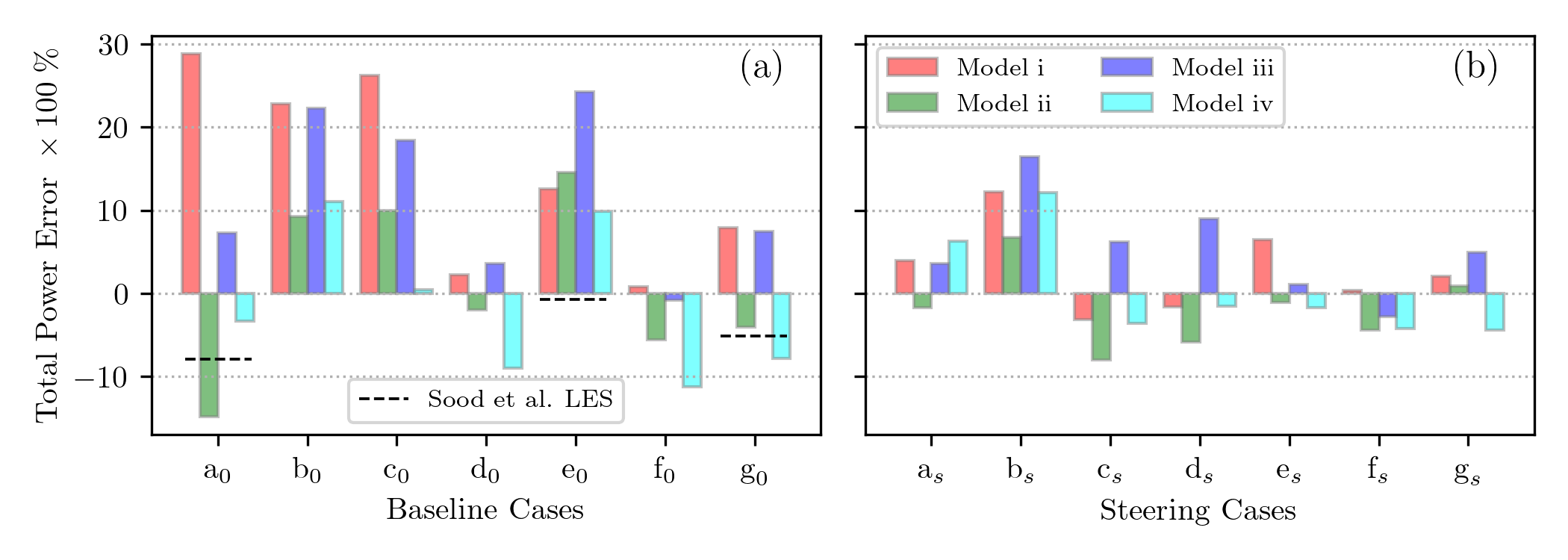}}
\caption{Error of the wind farm power for the (a) baseline cases and (b) steering cases. Dashed lines represent three LES cases from Sood et al \cite{sood2022comparison}. The legend in (b) applies to both subfigures. \label{fig_total_diff_p}}
\end{figure*}

It is also important to evaluate how well the models predict the total power output of the wind farm. Figure~\ref{fig_total_diff_p} shows the relative error in total power across all simulated cases compared with field measurements. A small total error does not necessarily imply accurate turbine-level predictions, as over- and underestimations may cancel out. For instance, in case~$\textrm{g}_s$, the MAE in Figure~\ref{fig_mae_p} is approximately 10\% for most models, yet the total power error remains within $\pm 5\%$ across all models. A similar pattern is observed in case~$\textrm{d}_0$, where models~i and~iii yield high MAE values but achieve relatively low total power errors. Conversely, in cases where errors are systematically biased, such as the overpredictions by model~i in case~$\textrm{a}_0$, the total error accumulates, resulting in a significant deviation from the measured power. Notably, the total power errors reported in the LES results of Sood et al~\cite{sood2022comparison} fall within a comparable range to those of the analytical models evaluated here.

Figure~\ref{fig_total_diff_p} shows that total power errors are often smaller in the steering cases than in the baseline set. This difference should be interpreted cautiously, as the two datasets differ in operating regime and atmospheric conditions; in particular, operation closer to rated wind speed can reduce the sensitivity of total power to inflow variability and may lead to partial error cancellation. Among the models, i and~iii tend to overpredict total power, with errors exceeding 20\% in some cases. In contrast, models~ii and~iv provide more consistent predictions, typically within a $\pm 10\%$ error range. However, model~iv shows notable deviations in cases~$\textrm{b}_0$ and~$\textrm{b}_s$, possibly due to inflow non-uniformity. Additional discrepancies in model~iv for cases~$\textrm{d}_0$ through~$\textrm{g}_0$ may be linked to spatial variations in the inflow across the upper region of the wind farm layout.

\section{Conclusions}\label{conclusions}

This study evaluated four combinations of steady-state analytical wake models using a comprehensive field dataset from the Lillgrund offshore wind farm, collected as part of the Horizon 2020 TotalControl campaign. The models, implemented in the LongSim software developed by DNV, incorporated distinct formulations for velocity deficit, added turbulence, wake superposition and deflection. Their performance was assessed under both baseline and active wake steering conditions using synchronous SCADA and LiDAR measurements. Overall, the models captured key trends in velocity deficit across a range of inflow directions and turbine alignments. In steering scenarios, they reproduced the general direction and extent of wake deflection caused by intentional yaw misalignment. The resulting power predictions agreed reasonably well with the field data, although accuracy tended to decrease with farm depth.

Case- and model-specific discrepancies persist, in part because the assumption of homogeneous inflow does not capture the spatial variability evident in the field data. Variations in upstream turbine power reflect this inflow heterogeneity, while errors tend to accumulate downstream due to model simplifications, including the omission of blockage effects. Model accuracy also declines in the near-wake region, where strong shear and steep velocity gradients pose challenges to analytical formulations. Differences in predictive performance highlight the sensitivity of wake expansion to turbulence intensity and other flow-dependent parameters. Across cases, the MAE of the normalised transect velocity deficit ranges from 7\% to 15\%. Models~ii and~iv yield more consistent predictions, with turbine-level power MAE values between 3\% and 10\%, and total wind farm power errors generally within $\pm 10\%$. These results are broadly comparable in magnitude to the LES benchmarks reported by Sood et al~\cite{sood2022comparison} for the three cases with similar wind-direction sectors.

Future improvements should focus on refining model representations of near-wake dynamics, incorporating more detailed inflow characterisation and improving wake predictions deeper in the farm, where cumulative effects are most pronounced. Extending these models to account for unsteady operating conditions, where sufficiently detailed inflow information is available, may further enhance their predictive reliability.

\bmsection*{Author contributions}

Diego Siguenza-Alvarado contributed to the conceptualisation and design of the study, developed the methodology, implemented the models and data processing routines, carried out the analysis and prepared the original manuscript draft. Matthew Harrison contributed to the methodology and software by supporting the use of LongSim and generating turbine performance data using the Bladed software developed by DNV. Majid Bastankhah initiated the research concept, supervised the work, validated the results and contributed to the manuscript revision. Ervin Bossanyi and Lars Landberg contributed to the study's conceptualisation and supervision; Ervin Bossanyi also supported model implementation and software development. All authors contributed to the interpretation of results and provided technical feedback during regular project discussions.

\bmsection*{Acknowledgments}
The authors gratefully acknowledge the support of the EPSRC Impact Acceleration Account at Durham University (EP/Y024052/1) for funding this project. Additional thanks are extended to Gunner C. Larsen and Elliot Simon for their valuable insights and for providing access to the Lillgrund dataset. We also thank Siemens Gamesa Renewable Energy for granting permission to use turbine data in this study.

\bmsection*{Funding}

This project was funded by the EPSRC Impact Acceleration Account at Durham University (EP/Y024052/1).

\bmsection*{Conflicts of interest}

The authors declare no conflicts of interest.

\bmsection*{Data Availability Statement}

The data that support the findings of this study are available on request from the corresponding author. The data are not publicly available due to privacy or ethical restrictions.

\bibliography{bib}


\bmsection*{Supporting information}

Additional supporting information may be found in the
online version of the article at the publisher’s website.

\appendix

\bmsection{Main modelling equations\label{app1}}
The following summarises the main equations for wake velocity deficit, added turbulence, wake superposition, and wake deflection for the four model combinations used in this work. Note that part of the nomenclature has been revised, relative to the original formulations, to ensure consistency across models. 

\bmsubsection{Model i Equations\label{app1.1}}

The velocity deficit in the Ainslie\ \cite{ainslie1988calculating} model takes the Gaussian form in cylindrical coordinates $(x,r)$, such that
\begin{equation}
    U_0 - U
    =
    \Delta\tilde{U}\,
    \exp\!\left(-3.56\left(\frac{r}{w_A}\right)^2\right),
\end{equation}
where $U_0$ is the free-stream streamwise velocity and $U$ is the streamwise wake velocity, while $\Delta\tilde{U}(x)$ is the centreline deficit, and $w_A$ denotes the Ainslie wake-width parameter. Both quantities are obtained implicitly from the numerical solution of Ainslie's eddy-viscosity model.

The streamwise and radial velocity $(V_r)$ fields are computed from the axisymmetric, steady Reynolds-averaged momentum equation:
\begin{equation}
U\frac{\partial U}{\partial x}
+
V_r\frac{\partial U}{\partial r}
=
-\frac{1}{r}
\frac{\partial}{\partial r}
\left(
r\,\overline{u'v'}
\right),
\label{eq:ainslie_pde}
\end{equation}
with the turbulent shear stress closed as $-\overline{u'v'}=\nu_T\,\frac{\partial U}{\partial r}$, where $\nu_T$ is the eddy viscosity, modified by Ruisi and Bossanyi\cite{ruisi2019engineering} such that
\begin{equation}
    \frac{\nu_T}{U_h\,d}
    =
    c\left(\frac{w_A}{d}\right)
    \left(\frac{\Delta\tilde{U}}{U_h}\right)
    +
    \frac{\kappa^2\,(z_h/d)}
    {\ln\!\left((z_h/z_0) + \Psi\right)\,\Phi},
\end{equation}
where $c$ is a DNV-calibrated mixing coefficient, $\kappa$ is the von~Kármán constant, $d$ is the rotor diameter, $z_h$ is the rotor hub-height, $U_h$ is the streamwise velocity at hub-height, and $z_0$ is the surface roughness length, inferred from the incoming shear. The stability correction functions $\Psi(z/L)$ and $\Phi(z/L)$ are defined by Ruisi and Bossanyi\cite{ruisi2019engineering} and depend on the Monin-Obukhov length $L$.

The quantities $\Delta\tilde{U}$ and $w_A$ are initially diagnosed using the formulations:
\begin{equation}
    \frac{\Delta{\tilde{U}}}{U_h}
    =
    c_T - 0.05 - (1.6\,c_T - 0.5)\,\left(\frac{\sigma_{u,\mathrm{local}}/U_{h,i}}{10}\right),
\end{equation}
and
\begin{equation}
    \frac{w_A}{d}
    =
    \sqrt{
    \frac{
        3.56\,c_T
    }{
        8\frac{\Delta{\tilde{U}}}{U_h}
        \left(1 - 0.5\frac{\Delta{\tilde{U}}}{U_h}\right)
    }}.
\end{equation}
Here $c_T$ is the turbine thrust coefficient, $U_{h,i}$ is the local velocity at hub-height, and $\sigma_{u,\mathrm{local}}$ is the local streamwise velocity standard deviation. The latter is obtained by $\sigma_{u,\mathrm{local}}=\sqrt{\sigma_u^2 + \sigma_{{u,\mathrm{added}}}^2}$, where $\sigma_u$ is the ambient velocity standard deviation referenced to the incoming hub-height wind speed, and $TI=\sigma_u/U_h$ denotes the ambient turbulence intensity. The wake-added contribution is combined at the standard-deviation level, while the resulting $\sigma_{u,\mathrm{local}}$ is normalised by the local hub-height velocity $U_{h,i}$ when used in the empirical deficit initialisation. The wake-added velocity standard deviation is defined according to Quarton and Ainslie\cite{quarton1990turbulence}, as modified by Hassan\ \cite{hassan1993wind}, considering the double Gaussian radial distribution proposed by Ishihara and Qian\cite{ishihara2018new} such that
\begin{equation}
    \sigma_{u,\mathrm{added}}
    =
    4.8\,\varphi\,
    U_{h,i}\,c_T^{0.7}\,TI^{0.68}
    (x/d)^{-0.57},
\end{equation}

where the distribution $\varphi$ is

\begin{equation}
    \varphi =
    k_1 \exp{\left(-\frac{(r-d/2)^2}{2 \sigma^2}\right)} +k_2 \exp{\left(-\frac{(r+d/2)^2}{2 \sigma^2}\right)},
\end{equation}

where $k_1$ and $k_2$ are shape functions defining the dual-Gaussian radial distribution, taken directly from the formulation of Ishihara and Qian\cite{ishihara2018new}.

For the Ainslie-based implementation considered here, the turbulence-spread parameter is taken as $\sigma=w_A/\sqrt{7.12}$, and the radial coordinate is defined as $r=\sqrt{(y-\delta)^2 + (z-z_h)^2}$, where the shift $y\rightarrow y-\delta$ accounts for wake-centre deflection $\delta$. The deflection is estimated using the model of Bastankhah and Port{\'e}-Agel\cite{bastankhah2016experimental}:
\begin{equation}
\frac{\delta}{d}
=
\theta_{\ell}\left(\frac{\ell}{d}\right)
+
\frac{\theta_{\ell}}{14.7\,k}
\sqrt{\frac{\cos\gamma}{c_T}}\,
\left( 2.9 + 1.3\sqrt{1 - c_T} \right)
\ln\!\left(
\frac{
\left(1.6 + \sqrt{c_T}\right)
\left( 1.6\sqrt{\tfrac{8\sigma_y\sigma_z}{d^2\cos\gamma}} - \sqrt{c_T} \right)
}{
\left(1.6 - \sqrt{c_T}\right)
\left( 1.6\sqrt{\tfrac{8\sigma_y\sigma_z}{d^2\cos\gamma}} + \sqrt{c_T} \right)
}
\right),
\label{eq:delta_2016}
\end{equation}
where $\gamma$ is the turbine yaw angle, $\theta_\ell$ is the initial skew angle in the near-wake region of length $\ell$, based on Coleman's potential-core formulation\cite{coleman1945evaluation}, and $\sigma_y$ and $\sigma_z$ are the characteristic wake widths in the lateral and vertical directions, respectively, taken as $\sigma_y=\sigma_z=\sigma$. Further details are provided in Bastankhah and Port{\'e}-Agel\cite{bastankhah2016experimental}.

Finally, wake superposition follows the “dominant wake” assumption of GL~2014\cite{gl2014theory}, in which the wake producing the largest deficit at a given turbine is taken as the governing contribution. For a given turbine $i$:
\begin{equation}
    \Delta U_i
    =
    \max\!\left(\Delta U_{1:i}\right).
\end{equation}

\bmsubsection{Model ii Equations\label{app1.2}}

The velocity deficit developed by Bastankhah et~al\ \cite{bastankhah2021analytical} incorporates the superposition of upwind turbine wakes by cumulatively solving the flow-governing equations:
\begin{equation}\label{eq:cum_def}
    U_0 - U_{i}
    =
    \sum_{j=1}^{i}
    \Delta\tilde{U}_j\,
    \exp\!\left(
      -\frac{
          \left( y - (y_j + \delta_j) \right)^2
          +
          \left( z - z_h \right)^2
      }{
          2\sigma_j^2
      }
    \right),
\end{equation}
where each Gaussian wake is laterally displaced by the deflection \(\delta_j\) of turbine \(j\). Here, $\sigma_j$ denotes the characteristic Gaussian wake width associated with turbine $j$. The corresponding centreline wake deficit of turbine \(i\) is given by
\begin{equation}\label{eq:Utilde_i}
    \frac{\Delta\tilde{U}_i}{U_h}
    =
    \left(
      1 -
      \sum_{j=1}^{i-1}
      \lambda_{ij}
      \frac{\Delta\tilde{U}_j}{U_h}
    \right)
    \left(
      1 -
      \sqrt{
        1 -
        \frac{
          \displaystyle c_{T,i}
          \left( \frac{{\langle U_{1:i-1} \rangle}_{(i,x_i)}}{U_h} \right)^2
        }{
          \displaystyle
          8\left( \frac{\sigma_i}{d} \right)^2
          \left(
            1
            -
            \sum_{j=1}^{i-1}
            \lambda_{ij}
            \frac{\Delta\tilde{U}_j}{U_h}
          \right)^2
        }
      }
    \right),
\end{equation}
where $\langle U_{1:i-1} \rangle_{(i,x_i)}$ is the rotor-averaged inflow streamwise velocity at the position $x_i$ of turbine $i$, and the dimensionless coefficient \(\lambda_{ij}\), which quantifies the contribution of turbine $j$ to the deficit of turbine $i$, is
\begin{equation}\label{eq:lambda}
    \lambda_{ij}
    =
    \frac{2\sigma_j^{2}}{\sigma_i^{2} + \sigma_j^{2}}
    \exp\!\left(
      -\frac{(y_i - y_j)^2}{2(\sigma_i^{2} + \sigma_j^{2})}
    \right)
    \exp\!\left(
      -\frac{(z_i - z_j)^2}{2(\sigma_i^{2} + \sigma_j^{2})}
    \right).
\end{equation}

The streamwise characteristic wake width is approximated as
\(
\sigma \approx \sqrt{\sigma_y \sigma_z},
\)
where the lateral and vertical wake widths are given by
\begin{equation}\label{eq:sigma}
\begin{cases}
    \sigma_y = k\left(x - \ell\right) + \sigma_\ell\,\cos\gamma, \\
    \sigma_z = k\left(x - \ell\right) + \sigma_\ell,
\end{cases}
\end{equation}

whereas the characteristic wake width at the onset of the far-wake region is
\begin{equation}\label{eq:sigma_ell}
 \sigma_\ell =
 \left(
 \frac{
 1 + \sqrt{1 - c_T \cos\gamma}
 }{
 8\left(1 + \sqrt{1 - c_T}\right)
 }
 \right)\,d ,
\end{equation}

while the wake growth parameter $k=0.3837\,TI + 0.003678$. 

Wake deflection \(\delta(x)\) is computed using Eq.~\ref{eq:delta_2016}, as in Model~i.

\bmsubsection{Model iii Equations \label{app1.3}}

Model~iii evaluates the velocity deficit by superposing the contributions of upwind turbines in a manner analogous to Model~ii, but additionally accounts for height-dependent wind veer by rotating the lateral coordinate. The wake velocity at turbine \(i\) is given by
\begin{equation}\label{eq:cum_def_veer}
    U_0 - U_{i}
    =
    \sum_{j=1}^{i}
    \Delta\tilde{U}_j\,
    \exp\!\left(
      -\frac{
          \left( y_v - (y_{v_j} + \delta_j) \right)^2
          +
          \left( z - z_h \right)^2
      }{
          2\sigma_j^2
      }
    \right),
\end{equation}
where the laterally rotated coordinate is defined as
\begin{equation}
y_v = -x\sin\!\big(\alpha(z)\big) + y\cos\!\big(\alpha(z)\big).
\end{equation}
with \(\alpha(z)\) denoting the wind-veer angle, representing the turning of the mean wind direction with height \(z\).

The corresponding centreline wake deficit of turbine \(i\) follows
Eq.~\ref{eq:Utilde_i}, as in Model~ii.

Model~iii departs from the wake-width formulations used previously by adopting a curled-vortex description\cite{bastankhah2022vortex,mohammadi2022curled}, which incorporates the asymmetric wake expansion induced by yaw misalignment and atmospheric veer.
Accordingly, the wake width is approximated as
\begin{equation}
{\sigma}^2(x)
\approx \left(k\,x + 0.4\,\tilde{\zeta}_0\right)
  \left(k\,x + 0.4\,\tilde{\zeta}_0 \cos\gamma\right)
\end{equation}
where the initial wake shape is
\begin{equation}
\tilde{\zeta}_0 \approx
R\sqrt{
\frac{1 + \sqrt{1 - c_T\cos^2\gamma}}
     {2\sqrt{1 - c_T\cos^2\gamma}}
},
\end{equation}
with $R$ denoting the rotor radius. Here, the wake growth parameter, $k=0.65\,TI$.

The wake deflection \(\delta(x)\) follows the analytical formulation:
\begin{equation}
\delta \approx
\left[
\frac{
(\pi-1)|\hat{t}|^3 + 2\sqrt{3}\,\pi^2 \hat{t}^2 + 48(\pi-1)^2|\hat{t}|
}{
2\pi(\pi-1)\hat{t}^2 + 4\sqrt{3}\,\pi^2 |\hat{t}| + 96(\pi-1)^2
}
\,\mathrm{sgn}(\hat{t})
\;-\;
\frac{2}{\pi}\,
\frac{\hat{t}}{\left[(z+z_h)/\tilde{\zeta}_0\right]^2 - 1}
\right]\tilde{\zeta}_0,
\end{equation}
where \(\hat{t}\) is a dimensionless time-like parameter governing wake evolution. It is approximated as
\begin{equation}
\hat{t}(x,z) \approx
-1.44\,\left(\frac{U_h}{u_*}\right)\,
\left(\frac{R}{\tilde{\zeta}_0}\right)\,
c_T \cos^2(\gamma)\sin(\gamma)
\left[
1 - \exp\!\left(
-0.35\,\frac{u_*}{U_0}\,
\frac{x}{R}
\right)
\right].
\end{equation}
with \(u_*\) denoting the friction velocity characterising the surface-layer shear stress, approximated from the wind-shear profile.

\bmsubsection{Model iv Equations \label{app1.4}}

This model extends the cumulative momentum-based framework of Bastankhah et~al~\cite{bastankhah2021analytical} to yawed wind turbines by incorporating the lateral momentum equation. The streamwise velocity deficit is modelled according to Eq.~\ref{eq:cum_def}. In the centreline wake deficit (Eq.~\ref{eq:Utilde_i}), the substitution $c_T \rightarrow c_T \cos\gamma$ is applied, while in the interaction coefficient $\lambda_{ij}$ (Eq.~\ref{eq:lambda}) the shift $y \rightarrow y + \delta$ is used when evaluating the interaction between an upstream turbine $j$ and a downstream turbine $i$.

The lateral wake velocity field is obtained cumulatively as
\begin{equation}\label{eq:asymmetric_gaussian}
    V_0 - V_i
    =
    \sum_{j=1}^{i}
    \Delta \tilde{V}_j
    \exp\!\left(
      -\frac{
          \left( y - (y_j + \epsilon_j) \right)^2
          +
          \left( z - z_h \right)^2
      }{
          2\tau_j^2
      }
    \right),
\end{equation}
where $V_0$ and $V_i$ denote the free-stream and wake-induced lateral velocity components, respectively. An asymmetric Gaussian profile is assumed to represent unequal lateral wake spreading. The characteristic lateral wake width $\tau$ is defined as
\begin{equation}\label{eq:tau_asymmetric}
\tau =
\begin{cases}
\tau_A \approx 0.75\,\sigma, & y\cdot \mathrm{sign} (\gamma) \ge \epsilon, \\[6pt]
\tau_B \approx 2\,\sigma, & y\cdot\mathrm{sign}(\gamma) < \epsilon .
\end{cases}
\end{equation}

The maximum induced lateral velocity is given by

\begin{equation}\label{eq:Vtilde_i}
    \frac{\Delta \tilde{V}_i}{U_h} = 
    \begin{cases}
    \displaystyle
    \frac{1}{4} c_{T,i} \, \left(\frac{\langle U_{1:i-1} \rangle_{(i,x_i)}}{U_h}\right) \, \sin{\gamma_i},
     & x_i < x \le x_i + \ell_i,
    \\[12pt]
    \displaystyle  
    \frac{
    \left( \frac{c_{T,i}}{8} \right)
    \left( \frac{{\langle U_{1:i-1} \rangle}_{(i,x_i)}}{U_h} \right)^2
    \sin\gamma_i
    +
    \frac{\Delta \tilde{U}_i}{U_h}
    \left( \frac{\sigma_i}{d} \right)^2
    \sum_{j=1}^{i-1}
    \frac{\Delta \tilde{V}_j}{U_h}\,\mu_{ij}
    }
    {
    \left( \frac{\tau_{A_i}}{d} \right)^2
    +
    \left( \frac{\tau_{B_i}}{d} \right)^2
    -
    \sum_{j=1}^{i-1}
    \frac{\Delta \tilde{U}_j}{U_h}
    \left( \frac{\sigma_j}{d} \right)^2
    \mu_{ji}
    -
    \frac{\Delta \tilde{U}_i}{U_h}
    \left( \frac{\sigma_i}{d} \right)^2
    \mu_{ii}
    } ,
     &  x \ge x_i + \ell_i.
    \end{cases}
\end{equation}

The dimensionless coefficient $\mu_{ij}$ (with indices adjusted for the cases $\mu_{ii}$ and $\mu_{ji}$) is defined as
\begin{equation}
\begin{aligned}
    \mu_{ij}
    &=
    \frac{\tau_{A_j}^2}{\sigma_i^2 + \tau_{A_j}^2}
    \exp\!\left(
    -\frac{\left( (y_i + \delta_i) - (y_j + \epsilon_j) \right)^2}
    {2\left( \sigma_i^2 + \tau_{A_j}^2 \right)}
    \right)
    \exp\!\left(
    -\frac{\left( z_i - z_j \right)^2}
    {2\left( \sigma_i^2 + \tau_{A_j}^2 \right)}
    \right)
    \\[4pt]
    &\quad +
    \frac{\tau_{B_j}^2}{\sigma_i^2 + \tau_{B_j}^2}
    \exp\!\left(
    -\frac{\left( (y_i + \delta_i) - (y_j + \epsilon_j) \right)^2}
    {2\left( \sigma_i^2 + \tau_{B_j}^2 \right)}
    \right)
    \exp\!\left(
    -\frac{\left( z_i - z_j \right)^2}
    {2\left( \sigma_i^2 + \tau_{B_j}^2 \right)}
    \right).
\end{aligned}
\end{equation}

The lateral wake centreline displacement $\epsilon$ is defined as
\begin{equation}\label{eq:epsilon_i}
    \epsilon =
    \begin{cases}
        \delta - \sigma, & \gamma > 0, \\
        \delta + \sigma, & \gamma < 0 .
    \end{cases}
\end{equation}

The streamwise characteristic wake widths are approximated as in Model~ii (see Equations~\ref{eq:sigma} and~\ref{eq:sigma_ell}), while the wake growth parameter $k=0.3837\,TI_{\mathrm{local}} + 0.003678$. The local turbulence intensity is obtained following the Crespo--Hern\'andez framework as a root-sum-square combination of the ambient and wake-added contributions, $TI_{\mathrm{local}}=\sqrt{TI^2+TI_{\mathrm{added}}^2}$.

The added turbulence generated within the wake is modelled using the updated formulation of Crespo and Hern\'andez~\cite{crespo1996turbulence} proposed by Zehtabiyan-Rezaie et~al~\cite{zehtabiyan2024wind}, accounting for wake overlap following Niayifar and Port{\'e}-Agel~\cite{niayifar2016analytical}. The maximum added turbulence at a downstream turbine $i$ induced by an upstream turbine $j$ is given by
\begin{equation}
    TI_{\mathrm{added},i}
    =
    \max_{j<i}\!\left[
    \left(\frac{A_j^*}{A_i}\right)
    (0.9\,a_j)^{0.8325}
    TI^{-0.0325}
    \left(\frac{x_i - x_j}{d}\right)^{-0.56}
    \right],
\end{equation}
where $A_j^*$ is the overlap area between the wake of turbine $j$ and the rotor area $A_i$ of turbine $i$. The axial induction factor is given by
\begin{equation}
    a = \tfrac{1}{2}\left(1 - \sqrt{1 - c_T}\right).
\end{equation}

Having established both the streamwise and lateral velocity fields, the wake deflection $\delta$ is obtained by solving an initial-value problem in the streamwise direction using an explicit Euler marching scheme:
\begin{equation}
    \delta(x')
    =
    \int_{0}^{x'}
    \frac{V_c}{U_c}\,\mathrm{d}x
    +
    \delta\big|_{x=0}.
\end{equation}
Here, $U_c$ and $V_c$ denote the streamwise and lateral velocity components evaluated at the wake centreline, respectively.

The velocity ratio evaluated at the wake centreline is given by
\begin{equation}\label{eq:theta_i}
  \frac{V_c}{U_c} =
  \begin{cases}
      \displaystyle
      \frac{
      \Delta \tilde{V}_i
      \exp\!\left( -\frac{\sigma_i^2}{2\tau_{A_i}^2} \right)
      }
      {
      U_{h,i} - \Delta \tilde{U}_i
      },
      & x_i < x \le x_i + \ell_i,
      \\[12pt]
      \displaystyle
      \frac{
      V_0(x,y_i+\delta_i,z_h)
      +
      \sum_{j=1}^{i}
      \Delta \tilde{V}_j
      \exp\!\left(
      -\frac{\left( (y_j + \epsilon_j) - (y_i + \delta_i) \right)^2}
      {2\tau_{A_j}^2}
      \right)
      }
      {
      U_{h,i}
      -
      \sum_{j=1}^{i}
      \Delta \tilde{U}_j
      \exp\!\left(
      -\frac{\left( (y_j + \delta_j) - (y_i + \delta_i) \right)^2}
      {2\sigma_j^2}
      \right)
      },
      & x > x_i + \ell_i .
  \end{cases}
\end{equation}

where $U_{h,i}$ denotes the local hub-height inflow velocity at the position of turbine $i$.

\end{document}